\documentclass[11pt,a4paper,numbers,sort&compress]{article}
\usepackage[utf8]{inputenc}
\pdfoutput=1
\usepackage{jheppub}

\usepackage{physics}
\usepackage{amsmath}
\usepackage{color}
\usepackage{xcolor}
\usepackage{cancel}
\usepackage{amsmath}
\usepackage{amssymb}
\usepackage{slashed}
\usepackage{cancel}
\usepackage[normalem]{ulem}
\newcommand{\lgrn}{\mathcal{L}}
\usepackage{tikz}

\usepackage{slashed}
\usepackage{cancel}
\usepackage[normalem]{ulem}

\newcommand{\nn}{\nonumber}
\newcommand{\fft}[2]{\frac{#1}{#2}}

\preprint{LCTP-24-07}

\title{Explicit Entropic Proofs of Irreversibility Theorems for Holographic RG Flows }

\author[a]{Evan Deddo}

\author[a]{James T. Liu}

\author[a,b]{Leopoldo A. Pando Zayas}

\author[a]{Robert J. Saskowski}

\emailAdd{evdedd@umich.edu, jimliu@umich.edu, lpandoz@umich.edu, rsaskows@umich.edu}

\affiliation[a]{Leinweber Center for Theoretical Physics, 
University of Michigan, Ann Arbor, MI 48109, USA}

\affiliation[b]{The Abdus Salam International Centre for Theoretical Physics, 34014 Trieste, Italy}

\abstract{We revisit the existence of monotonic quantities along renormalization group flows using only the Null Energy Condition and the Ryu-Takayanagi formula for the entanglement entropy of field theories with anti-de Sitter gravity duals. In particular, we consider flows within the same dimension and holographically reprove the $c$-, $F$-, and $a$-theorems in dimensions two, three, and four. We focus on the family of maximally spherical entangling surfaces, define a quasi-constant of motion corresponding to the breaking of conformal invariance, and use a properly defined distance between minimal surfaces to construct a holographic $c$-function that is monotonic along the flow. We then apply our method to the case of flows across dimensions: There, we reprove the monotonicity of flows from $\mathrm{AdS}_{D+1}$ to $\mathrm{AdS}_3$ and prove the novel case of flows from $\mathrm{AdS}_5$ to $\mathrm{AdS}_4$.}

\date{\today}

\begin{document}

\maketitle
\section{Introduction  and summary}
A central organizing principle in quantum field theory is that of renormalization group (RG) flows. In general, we expect the process of coarse-graining along flows to result in a reduction in degrees of freedom, generally measured by a suitable choice of $c$-function. Establishing the irreversibility of RG flows involves distinct field-theoretic techniques in different spacetime dimensions. For example, the $c$-theorem, establishing the irreversibility of flows in two dimensions, exploits properties of the correlators of two stress-energy tensors~\cite{Zamolodchikov:1986gt}, while the four-dimensional $a$-theorem is proved non-perturbatively with the help of a certain four-point amplitude~\cite{Komargodski:2011vj}, and progress on the six-dimensional $a$-theorem has focused on four- and six-point amplitudes~\cite{Elvang:2012st,Cordova:2015vwa,Cordova:2015fha,Heckman:2021nwg}. Entropic approaches first emerged in (re)proving the $c$-theorem in~\cite{Casini:2004bw,Casini:2006es} and resulted in the first proof of the three-dimensional $F$-theorem \cite{Casini:2012ei}; such approaches have proven to be extremely successful in providing a general framework, and a unified proof for spacetime dimensions two, three, and four was presented in~\cite{Casini:2017vbe}. The latter approach is rooted in the strong subadditivity (SSA) property of the relative entanglement entropy  (EE). The authors of~\cite{Casini:2017vbe} used SSA to show that for a field theory in $d$ dimensions, a suitably subtracted EE satisfies the inequality
\begin{align}
	\label{Eq:MainSubAd}
	R \partial^2_{R}\, \Delta S_\mathrm{EE}(R)-(d-3) \partial_R\,\Delta  S_\mathrm{EE}(R)&\leq 0,
\end{align}
where $R$ is the radius of a spherical entangling region. Here,
\begin{equation}
    \Delta S_\text{EE}(R)\equiv S_\mathrm{EE}(R)-S_\mathrm{UV},\label{eq:sub}
\end{equation}
where $S_\mathrm{UV}$ is the EE of the UV fixed point, the subtraction of which regulates the divergences. The above result, therefore,  establishes a monotonically non-increasing function, 
\begin{equation}
    c_\text{CTT}(R)=R \partial_R\Delta S(R) - (d-2) \Delta S(R),\label{eq:cCTT}
\end{equation}
along RG flows. In  $d\leq 3$, this function $c_\text{CTT}(R)$ interpolates between the UV and IR central charges or free energies.\footnote{As an abuse of language, we will generally refer to both of these as central charges, even in odd dimensions where the Weyl anomaly vanishes.} This is no longer the case in higher dimensions, but in $d=4$, the inequality is still strong enough to prove the $a$-theorem. The connection between the amplitude and EE approaches to establishing the irreversibility of RG flows has remained largely open.

Recently, there has been some work on connecting the correlators that arise in proving $c$-theorems with the Averaged Null Energy Condition (ANEC), which can be derived both holographically~\cite{Kelly:2014mra} and field-theoretically ~\cite{Klinkhammer:1991ki,Faulkner:2016mzt,Hartman:2016lgu,Kravchuk:2018htv}. In CFT, the ANEC provides several constraints on the coupling constants and anomaly coefficients~\cite{Hofman:2008ar,Hofman:2009ug} and results in the 2d $c$-theorem and the 4d $a$-theorem being related to particular three-point correlators involving the averaged null energy~\cite{Hartman:2023qdn,Hartman:2023ccw}. Since the averaged null energy condition is related to the modular Hamiltonian \cite{Faulkner:2016mzt}, this proof may be rephrased in terms of relative entropy, which was also used in \cite{Casini:2016udt} to prove the $c$-theorem. A second, related condition is the Quantum Null Energy Condition (QNEC) \cite{Bousso:2015mna}, which has been proved field-theoretically in~\cite{Bousso:2015wca,Balakrishnan:2017bjg,Malik:2019dpg,Balakrishnan:2019gxl,Kudler-Flam:2023hkl} and holographically in~\cite{Koeller:2015qmn,Leichenauer:2018obf}, and is implied by the ANEC \cite{Ceyhan:2018zfg}. In particular, \cite{Ecker:2020gnw} showed that the inequality that arises in proving the 2d $c$-theorem is precisely the QNEC, and \cite{Casini:2023kyj} used the QNEC to reprove \eqref{Eq:MainSubAd} and extend it to defect RG flows.

Much has been accomplished specifically using holographic methods. In particular, one approach is to use the NEC to directly construct a local holographic $c$-function that monotonically interpolates between the UV and IR central charges (see \emph{e.g.}~\cite{Girardello:1998pd,Freedman:1999gp,Myers:2010tj,Myers:2010xs,Anber:2008js,Sinha:2010ai,Gullu:2010pc,Gullu:2010st,Sinha:2010pm,Oliva:2010eb,Myers:2010ru,Oliva:2010zd,Liu:2010xc,Liu:2011iia,Alkac:2018whk,Ghodsi:2019xrx,Anastasiou:2021jcv,Ghodsi:2021xrb,Alkac:2022zda}). This method is powerful and leads to proofs in arbitrary dimensions, but the resulting $c$-function lacks a clear field-theoretic interpretation away from the fixed points. This is because taking an arbitrary slice of AdS away from the boundary does not necessarily have the interpretation of being dual to a QFT. In particular, the bulk (A)NEC is not holographically dual to the boundary ANEC \cite{Hartman:2022njz}.

A complementary approach is to use holographic EE, and some progress has been made in this direction~\cite{Casini:2011kv,Albash:2011nq,Myers:2012ed,Liu:2013una,Casini:2015ffa,Park:2018ebm,Daguerre:2022uxt}.
As expected of an EE, the holographic EE has been shown to satisfy SSA \cite{Wall:2012uf}, and so one may use the same proof on the holographic EE as for the field-theoretic EE, although this is quite circuitous. We will mainly be interested in genuinely holographic proofs. In particular, \cite{Myers:2012ed} considered a strip-shaped entangling surface and used the integral of motion and the NEC to show monotonicity. We expect a spherical entangling region to pick out the $A$-type central charge, whereas the strip selects some linear combination of central charges. In $d=2$, the strip is a sphere, and so \cite{Myers:2012ed} provides a holographic $c$-function that yields the 2d $c$-theorem, but for $d>2$, the strip does not yield the $A$-type central charge. The two-dimensional $c$-theorem was further investigated in \cite{Casini:2015ffa,Park:2018ebm}. More generally, the difficulty with spherical entangling regions is that there is no integral of motion for dimensions $d>2$, which makes them computationally difficult compared with a strip. 

More recently, \cite{Daguerre:2022uxt} used the Hamilton-Jacobi equations associated to the Ryu-Takayanagi surface \cite{Ryu:2006bv,Ryu:2006ef} to derive a sequence of equations for the $\mu_{d-2k}$ coefficients in the IR expansion of the entanglement entropy of the sphere
\begin{equation}
    S_\mathrm{EE}(R)=\mu_{d-2}(\epsilon)R^{d-2}+\mu_{d-4}(\epsilon)R^{d-4}+\cdots+\begin{cases}
        (-1)^{\tfrac{d}{2}+1}4A\log\frac{R}{\epsilon}&d\text{ even}\\
        (-1)^{\tfrac{d-1}{2}}F&d\text{ odd}
    \end{cases},
\end{equation}
where $\epsilon$ is a UV cutoff. When compared to the non-flowing theory, this allowed the authors to compute the differences $\Delta\mu_{d-2k}$, which resulted in a holographic proof of the $c$-, $F$-, and $a$-theorems as a comparison between the UV and IR central charges, but they did not construct a $c$-function along the flow. Hence, the first goal of the present work will be to construct such a holographic $c$-function as in \eqref{eq:cCTT} that continuously measures degrees of freedom along the RG flow and to demonstrate its monotonicity.

Additionally, there has been some work on flows across dimensions recently, which may be viewed as flows that start at a $D$-dimensional UV fixed point, undergo compactification onto a $(D-d)$-dimensional manifold along the flow, and end at a $d$-dimensional IR fixed point. This is quite difficult to study directly in field theory but is holographically dual to a gravitational solution interpolating between $\mathrm{AdS}_{D+1}$ and $\mathrm{AdS}_{d+1}$. There are many examples of supergravity solutions holographically dual to RG flows interpolating between CFTs of different dimensions~\cite{Maldacena:2000mw,Acharya:2000mu,Gauntlett:2000ng,Gauntlett:2001jj,Gauntlett:2001qs,Benini:2013cda,Benini:2015bwz,Bobev:2017uzs}, and some candidate $c$-functions for such flows were studied in~\cite{Macpherson:2014eza,Bea:2015fja,Legramandi:2021aqv}. More recently, explicit $c$-functions were constructed in~\cite{GonzalezLezcano:2022mcd,Deddo:2023pid} and further explored in~\cite{Deddo:2022wxj,de-la-Cruz-Moreno:2023mew}. In particular, \cite{GonzalezLezcano:2022mcd,Deddo:2023pid} proved the monotonicity of flows to $\mathrm{AdS}_3$, but it is difficult to push the analysis further due to the lack of a first integral for the surface minimization equation. A second goal of this manuscript is to generalize our results for flows within the same dimension to those across dimensions.

In this work, we provide a unified proof, valid in various dimensions, using only the Null Energy Condition (NEC) and the assumption that for field theories with a gravity dual the Ryu-Takayanagi (RT) formula~\cite{Ryu:2006bv,Ryu:2006ef} determines the entanglement entropy of a region. We start by considering flows within the same dimension, for which we will prove the inequality \eqref{Eq:MainSubAd} holographically.  These flows correspond to a domain wall solution
\begin{equation}
    \dd s^2=\frac{1}{h(z)^2}\qty(-\dd t^2+\dd z^2+\dd r^2+r^2\dd\Omega_{d-2}^2),
\end{equation}
where $\dd\Omega_{d-2}^2$ is the line element on the unit sphere $S^{d-2}$ and $h(z)$ asymptotes to $z/L_\mathrm{UV}$ in the UV and $z/L_\mathrm{IR}$ in the IR, where $L_\mathrm{UV}$ and $L_\mathrm{IR}$ correspond to the UV and IR AdS radii, respectively.  Using the RT prescription, we compute the entanglement entropy as the area of the minimal surface wrapping $S^{d-2}$ with radius $R$ on the conformal boundary and use this along with the NEC to obtain the inequality \eqref{Eq:MainSubAd}.  Due to the spherical symmetry of the geometry, we may assume that the RT surface has a profile $r(z)$ with $r(0)=R$ independent of the angular coordinates.

Motivated by the presence of $R$ derivatives in \eqref{Eq:MainSubAd}, we introduce a continuous family of such surfaces parameterized by $R$. Differentiating the equations of motion for $r(z)$ yields a linear equation for $\dd r/\dd R$. The key step is then to introduce a more suitable measurement of the separation between infinitesimally close surfaces,
\begin{equation}
    v(z)\equiv\frac{r(z)}{R}\dv{r}{R},
\label{eq:v(z)def}
\end{equation}
which we will show satisfies $v'(z)\ge 0$. The proof then reduces to the observation that
\begin{equation}
    \dv{}{R}\qty(\frac{\Delta S'_\mathrm{EE}(R)}{R^{d-3}})\sim -v'(z),
\end{equation}
so long as the conformal dimension of the perturbing coupling satisfies
\begin{equation}\label{eq:delbound}
    \Delta_+\ge \frac{d-4}{2}.
\end{equation}
Given that $\Delta_+>0$, this is only an additional condition when $d\ge 5$. In particular, \eqref{eq:delbound} recovers the results of \cite{Casini:2017vbe} and provides a holographic, monotonic $c$-function, proving the $c$-, $F$-, and $a$-theorems.

We then consider the case of flows across dimensions, which corresponds to a metric of the form 
\begin{equation}
    \dd s^2=\frac{1}{h(z)^2}\qty(-\dd t^2+\dd z^2+\dd r^2+r^2\dd\Omega_{d-2}^2)+\frac{1}{j(z)^2}\dd s^2_{M_{D-d}},
\end{equation}
subject to the asymptotics
\begin{align}
    h(z)\overset{z\to 0}{\sim}\frac{z}{L_\mathrm{UV}},&\qquad h(z)\overset{z\to\infty}{\sim}\frac{z}{L_\mathrm{IR}},\nonumber\\
    j(z)\overset{z\to 0}{\sim}\frac{z}{L_\mathrm{UV}},&\qquad j(z)\overset{z\to\infty}{\sim}j_\mathrm{IR}.
\end{align}
We again make use of the RT formula to construct a central charge for entangling regions that wrap the internal manifold. We begin by reproving the case of flows from $\mathrm{AdS}_{D+1}$ to $\mathrm{AdS}_3$, which was originally proved in \cite{GonzalezLezcano:2022mcd}, in our new formalism. We then attempt to extend the proof to general flows to $\mathrm{AdS}_{d+1}$. We find that additional divergences arise that hamper a proof in general dimensions, but we do prove the monotonicity of flows from $\mathrm{AdS}_5$ to $\mathrm{AdS}_4$.

The rest of this paper is organized as follows. In Section \ref{sec:sameDim}, we consider flows within the same dimension and prove the $c$-, $F$-, and $a$-theorems using holographic EE and the NEC. In Section \ref{sec:acrossDims}, we extend our results to the case of flows across dimensions. Finally, we conclude in Section \ref{sec:discussion} and discuss some future directions.

\section{Flows within the same dimension}\label{sec:sameDim}

We start with holographic flows in the same dimension.  In order to prove the inequality, \eqref{Eq:MainSubAd}, for the subtracted entanglement entropy $\Delta S_{\mathrm{EE}}(R)$, we first need to understand the properties of holographic domain wall flow.  This will be constrained by the bulk NEC.  Having understood the background, we then probe its geometry by constructing the RT surface with profile $r(z)$ and computing its corresponding entropy function $S(R)$.  Because of its property as a minimal surface, we can obtain its derivative, $S'(R)$, as a local boundary expression.

In principle, knowledge of $S'(R)$ is all we need to construct the $c_{\mathrm{CTT}}(R)$ function of (\ref{eq:cCTT}).  However, proof of its monoticity requires a second derivative with respect to $R$, as in (\ref{Eq:MainSubAd}), and this requires additional work.  Essentially, what we are looking for is the change in the surface (and its corresponding area) as we vary the boundary radius $R$.  We find that this is well captured by the separation function $v(z)$ as defined in (\ref{eq:v(z)def}).  By studying the breaking of conformal symmetry as one moves away from the UV boundary, we are able to show that the separation of two nearby minimal surfaces increases as one moves into the interior of AdS$_{d+1}$ in the sense that $v'(z)\ge0$.  Finally, we complete the proof of monotonicity of the $c$-function, $\partial_Rc_{\mathrm{CTT}}(R)\le0$, by relating this derivative to the behavior of $v'(z)$ near the UV boundary.

In summary, the holographic proof proceeds as follows:
\begin{enumerate}
    \item We first obtain a condition on the domain wall metric warp factor, $h''(z)\ge 0$ by imposing the NEC in the bulk.  This is simply an application of the usual local holographic $c$-theorem \cite{Freedman:1999gp}.
    \item We introduce a profile $r(z)$ for the RT surface, and find an expression for $S'(R)$ via the Hamilton-Jacobi equation. This gives an expression for $S'$ in terms of local boundary values of fields.
    \item We construct a monotonically non-increasing function $\mathcal C(z)$ built from the metric function $h(z)$ and the surface profile $r(z)$ that measures the breaking of conformal symmetry. Moreover, we show that $\mathcal C\ge 0$ everywhere along the surface.
    \item We define the separation function $v(z)$ according to (\ref{eq:v(z)def}) and prove that the non-negativity of $\mathcal C(z)$ and the NEC implies that $v'(z)\ge 0$.
    \item Finally, we show that $\partial_R\,c_{\mathrm{CTT}}(R)$ is proportional to the leading term in $-v'(z)$ near the boundary.  Since $v'(z)\ge0$ everywhere along the flow, this immediately leads to the entropic $c$-theorem, $\partial_R\,c_{\mathrm{CTT}}(R)\le0$.
\end{enumerate}

\subsection{AdS/CFT and domain wall flows}
The AdS/CFT correspondence \cite{Maldacena:1997re} posits an equivalence between certain field theories and theories of gravity, leading to a geometrization of various field theory concepts. For example, conformal symmetry in the field theory is mapped to the isometries of the gravity background. We consider a metric of the form
\begin{equation}\label{eq:metric}
    \dd s^2=\fft1{h(z)^2}\qty(\eta_{\mu\nu}\dd x^\mu \dd x^\nu+\dd z^2)=\frac{1}{h(z)^2}\qty(-\dd t^2+\dd z^2+\dd r^2+r^2\dd\Omega_{d-2}^2),
\end{equation}
where $\eta_{\mu\nu}$ is the Minkowski metric and $\dd\Omega_{d-2}^2$ is the line element on $S^{d-2}$. This spacetime may be viewed as a domain wall solution interpolating between a UV and an IR $\mathrm{AdS}_{d+1}$, so long as we have appropriate asymptotics of the form
\begin{align}
    h(z)\overset{z\to 0}{\sim}\frac{z}{L_\mathrm{UV}},&\qquad h(z)\overset{z\to\infty}{\sim}\frac{z}{L_\mathrm{IR}}.
\end{align}
Under the AdS/CFT dictionary, we may view this as an RG flow between two $d$-dimensional CFTs~\cite{Girardello:1998pd,Freedman:1999gp}. In the rest of this manuscript, we will set $L_\mathrm{UV}=1$ for convenience.

We now constrain the domain wall flow by imposing the null energy condition in the bulk.  Given a future-directed null vector $\xi$, the NEC requires that
\begin{equation}
    T_{MN}\xi^M\xi^N\ge0,
\end{equation}
where $T_{MN}$ is the stress-energy tensor. Making use of the Einstein equations, the stress-energy tensor may be exchanged for the Ricci tensor
\begin{equation}
    \hat R_{MN}\xi^M\xi^N\ge0.
\end{equation}
The metric \eqref{eq:metric} has curvature tensor components
\begin{equation}
    \hat R^\mu{}_{\nu\lambda\sigma}=-\frac{h'^2}{h^2}(\delta^\mu_\lambda \eta_{\nu\sigma}-\delta^\mu_\sigma \eta_{\nu\lambda}),\qquad \hat R^\mu{}_{z\nu z}=\frac{1}{h^2}(hh''-h'^2)\delta^\mu_\nu,
\end{equation}
and
\begin{equation}
    \hat R_{\mu\nu}=\frac{1}{h^2}(hh''-dh'^2)\eta_{\mu\nu},\qquad \hat R_{zz}=\frac{d}{h^2}(hh''-h'^2).
\end{equation}
Picking $\xi=\partial_z\pm\partial_t$, the NEC then implies
\begin{equation}\label{eq:NEC}
    h''(z)\geq 0.
\end{equation}
Note that this will be the only non-trivial NEC due to the planar symmetry of the domain
wall. The main consequence is that the slope of $h$ must be larger in the IR than in the UV, and so the effective radius 
\begin{equation}
    L_\mathrm{eff}(z)=\frac{1}{h'(z)},
\end{equation}
decreases along the flow. Since the AdS radius is a proxy for the number of degrees of freedom, one may naturally write down an (unnormalized) $c$-function
\begin{equation}
    c_\mathrm{NEC}(z)=L_\mathrm{eff}(z)^{d-1},
\end{equation}
and obtain a holographic $c$-theorem \cite{Freedman:1999gp,Myers:2010xs}. However, as already emphasized, this $c$-function only has a rigorous field-theoretic interpretation at the fixed points of the flow.

\subsection{Entanglement entropy and the RT surface}
\begin{figure}[t]
    \centering

\tikzset{every picture/.style={line width=0.75pt}} 

\begin{tikzpicture}[x=0.75pt,y=0.75pt,yscale=-1.3,xscale=1.3]

\draw    (291.5,300.1) -- (488,300) ;
\draw [shift={(490,300)}, rotate = 179.97] [color={rgb, 255:red, 0; green, 0; blue, 0 }  ][line width=0.75]    (10.93,-3.29) .. controls (6.95,-1.4) and (3.31,-0.3) .. (0,0) .. controls (3.31,0.3) and (6.95,1.4) .. (10.93,3.29)   ;
\draw    (291.5,300.1) -- (290.02,154.6) ;
\draw [shift={(290,152.6)}, rotate = 89.42] [color={rgb, 255:red, 0; green, 0; blue, 0 }  ][line width=0.75]    (10.93,-3.29) .. controls (6.95,-1.4) and (3.31,-0.3) .. (0,0) .. controls (3.31,0.3) and (6.95,1.4) .. (10.93,3.29)   ;
\draw  [color={rgb, 255:red, 126; green, 211; blue, 33 }  ,draw opacity=1 ][fill={rgb, 255:red, 126; green, 211; blue, 33 }  ,fill opacity=1 ] (288,176) .. controls (288,174.9) and (288.9,174) .. (290,174) .. controls (291.1,174) and (292,174.9) .. (292,176) .. controls (292,177.1) and (291.1,178) .. (290,178) .. controls (288.9,178) and (288,177.1) .. (288,176) -- cycle ;
\draw    (291.5,300.1) -- (210.48,373.26) ;
\draw [shift={(209,374.6)}, rotate = 317.92] [color={rgb, 255:red, 0; green, 0; blue, 0 }  ][line width=0.75]    (10.93,-3.29) .. controls (6.95,-1.4) and (3.31,-0.3) .. (0,0) .. controls (3.31,0.3) and (6.95,1.4) .. (10.93,3.29)   ;
\draw  [color={rgb, 255:red, 74; green, 144; blue, 226 }  ,draw opacity=1 ][fill={rgb, 255:red, 74; green, 144; blue, 226 }  ,fill opacity=0.24 ] (156.38,300.1) .. controls (156.38,282.29) and (216.87,267.85) .. (291.5,267.85) .. controls (366.13,267.85) and (426.63,282.29) .. (426.63,300.1) .. controls (426.63,317.91) and (366.13,332.35) .. (291.5,332.35) .. controls (216.87,332.35) and (156.38,317.91) .. (156.38,300.1) -- cycle ;
\draw  [draw opacity=0] (156.37,298.88) .. controls (156.35,230.75) and (216.89,175.51) .. (291.57,175.49) .. controls (366.26,175.48) and (426.81,230.7) .. (426.82,298.83) -- (291.59,298.85) -- cycle ; \draw  [color={rgb, 255:red, 208; green, 2; blue, 27 }  ,draw opacity=1 ] (156.37,298.88) .. controls (156.35,230.75) and (216.89,175.51) .. (291.57,175.49) .. controls (366.26,175.48) and (426.81,230.7) .. (426.82,298.83) ;  
\draw [color={rgb, 255:red, 144; green, 19; blue, 254 }  ,draw opacity=1 ]   (290.75,226.35) -- (400,226.6) ;
\draw [shift={(402,226.6)}, rotate = 180.13] [color={rgb, 255:red, 144; green, 19; blue, 254 }  ,draw opacity=1 ][line width=0.75]    (10.93,-3.29) .. controls (6.95,-1.4) and (3.31,-0.3) .. (0,0) .. controls (3.31,0.3) and (6.95,1.4) .. (10.93,3.29)   ;
\draw  [draw opacity=0] (416.68,328.27) .. controls (396.43,340.64) and (348.04,349.35) .. (291.56,349.35) .. controls (291.12,349.35) and (290.68,349.35) .. (290.24,349.35) -- (291.56,315.31) -- cycle ; \draw [color={rgb, 255:red, 65; green, 117; blue, 5 }  ,draw opacity=1 ]   (416.68,328.27) .. controls (396.43,340.64) and (348.04,349.35) .. (291.56,349.35) ; \draw [shift={(290.24,349.35)}, rotate = 359.51] [color={rgb, 255:red, 65; green, 117; blue, 5 }  ,draw opacity=1 ][line width=0.75]    (10.93,-4.9) .. controls (6.95,-2.3) and (3.31,-0.67) .. (0,0) .. controls (3.31,0.67) and (6.95,2.3) .. (10.93,4.9)   ; 
\draw  [color={rgb, 255:red, 126; green, 211; blue, 33 }  ,draw opacity=1 ][fill={rgb, 255:red, 126; green, 211; blue, 33 }  ,fill opacity=1 ] (288.4,174.8) .. controls (288.4,173.81) and (289.21,173) .. (290.2,173) .. controls (291.19,173) and (292,173.81) .. (292,174.8) .. controls (292,175.79) and (291.19,176.6) .. (290.2,176.6) .. controls (289.21,176.6) and (288.4,175.79) .. (288.4,174.8) -- cycle ;
\draw  [draw opacity=0] (426.71,300.06) .. controls (426.71,300.06) and (426.71,300.06) .. (426.71,300.06) .. controls (426.71,300.06) and (426.71,300.06) .. (426.71,300.06) .. controls (426.71,317.9) and (366.22,332.35) .. (291.59,332.35) .. controls (216.97,332.35) and (156.48,317.9) .. (156.48,300.06) -- (291.59,300.06) -- cycle ; \draw  [color={rgb, 255:red, 208; green, 2; blue, 27 }  ,draw opacity=1 ] (426.71,300.06) .. controls (426.71,300.06) and (426.71,300.06) .. (426.71,300.06) .. controls (426.71,300.06) and (426.71,300.06) .. (426.71,300.06) .. controls (426.71,317.9) and (366.22,332.35) .. (291.59,332.35) .. controls (216.97,332.35) and (156.48,317.9) .. (156.48,300.06) ;  
\draw  [draw opacity=0] (426,285.6) .. controls (426,285.6) and (426,285.6) .. (426,285.6) .. controls (426,300.51) and (365.87,312.6) .. (291.7,312.6) .. controls (217.52,312.6) and (157.39,300.51) .. (157.39,285.6) -- (291.7,285.6) -- cycle ; \draw  [color={rgb, 255:red, 80; green, 227; blue, 194 }  ,draw opacity=1 ] (426,285.6) .. controls (426,285.6) and (426,285.6) .. (426,285.6) .. controls (426,300.51) and (365.87,312.6) .. (291.7,312.6) .. controls (217.52,312.6) and (157.39,300.51) .. (157.39,285.6) ;  
\draw  [draw opacity=0][dash pattern={on 4.5pt off 4.5pt}] (425.59,284.04) .. controls (420,268.12) and (362.17,255.68) .. (291.68,255.76) .. controls (217.49,255.85) and (157.37,269.79) .. (157.39,286.89) -- (291.71,286.72) -- cycle ; \draw  [color={rgb, 255:red, 80; green, 227; blue, 194 }  ,draw opacity=1 ][dash pattern={on 4.5pt off 4.5pt}] (425.59,284.04) .. controls (420,268.12) and (362.17,255.68) .. (291.68,255.76) .. controls (217.49,255.85) and (157.37,269.79) .. (157.39,286.89) ;  
\draw [color={rgb, 255:red, 80; green, 227; blue, 194 }  ,draw opacity=1 ]   (434,284.6) -- (434,298.6) ;
\draw [shift={(434,298.6)}, rotate = 270] [color={rgb, 255:red, 80; green, 227; blue, 194 }  ,draw opacity=1 ][line width=0.75]    (0,5.59) -- (0,-5.59)   ;
\draw [shift={(434,284.6)}, rotate = 270] [color={rgb, 255:red, 80; green, 227; blue, 194 }  ,draw opacity=1 ][line width=0.75]    (0,5.59) -- (0,-5.59)   ;
\draw [color={rgb, 255:red, 245; green, 166; blue, 35 }  ,draw opacity=1 ]   (291.7,285.6) -- (424,285.6) ;
\draw [shift={(426,285.6)}, rotate = 180] [color={rgb, 255:red, 245; green, 166; blue, 35 }  ,draw opacity=1 ][line width=0.75]    (10.93,-3.29) .. controls (6.95,-1.4) and (3.31,-0.3) .. (0,0) .. controls (3.31,0.3) and (6.95,1.4) .. (10.93,3.29)   ;

\draw (297,157.4) node [anchor=north west][inner sep=0.75pt]  [color={rgb, 255:red, 126; green, 211; blue, 33 }  ,opacity=1 ]  {$z_{0}$};
\draw (285,134.4) node [anchor=north west][inner sep=0.75pt]    {$z$};
\draw (390,184.4) node [anchor=north west][inner sep=0.75pt]  [color={rgb, 255:red, 208; green, 2; blue, 27 }  ,opacity=1 ]  {$\gamma $};
\draw (143,321.4) node [anchor=north west][inner sep=0.75pt]  [color={rgb, 255:red, 74; green, 144; blue, 226 }  ,opacity=1 ]  {$M$};
\draw (414,217.4) node [anchor=north west][inner sep=0.75pt]  [color={rgb, 255:red, 144; green, 19; blue, 254 }  ,opacity=1 ]  {$r( z)$};
\draw (365,349.4) node [anchor=north west][inner sep=0.75pt]  [color={rgb, 255:red, 65; green, 117; blue, 5 }  ,opacity=1 ]  {$\Omega _{d-2}$};
\draw (442,282.4) node [anchor=north west][inner sep=0.75pt]  [color={rgb, 255:red, 80; green, 227; blue, 194 }  ,opacity=1 ]  {$\epsilon $};
\draw (428,265.4) node [anchor=north west][inner sep=0.75pt]  [color={rgb, 255:red, 245; green, 166; blue, 35 }  ,opacity=1 ]  {$R_{c}$};

\end{tikzpicture}

    \caption{Representation of the minimal surface $\gamma$ homologous to a spherical entangling region $M$ of radius $R$. Here, $\Omega_{d-2}$ represents all the angular directions on the $S^{d-2}$, and $\epsilon$ indicates the holographic cutoff, where the surface is anchored such that $z(R_c)=\epsilon$.}
    \label{fig:RTsurf}
\end{figure}
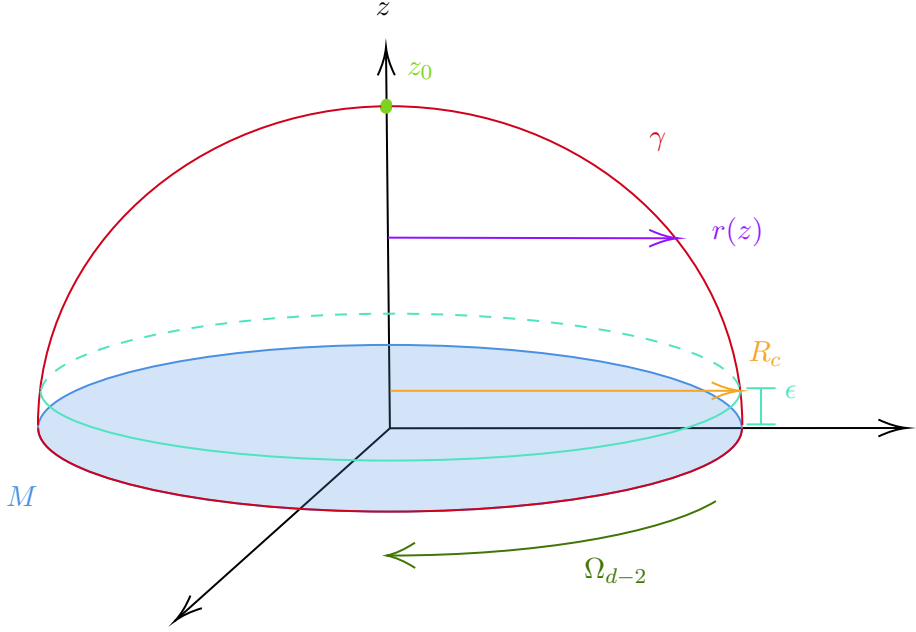

The Ryu-Takayanagi formula \cite{Ryu:2006bv,Ryu:2006ef} allows one to compute the entanglement entropy of a region, $M$, in field theory as the area of the bulk surface homologous to $M$. For the case of a  minimal surface $\gamma$ anchored on the maximally spherical region of radius $R$ on the boundary and parametrized by $r(z)$, using the metric \eqref{eq:metric}, we find that
\begin{equation}
    {\cal A}(\gamma) = \mathrm{Vol}(S^{d-2})\int_\epsilon^{z_0}\dd z\;\frac{r(z)^{d-2}}{h(z)^{d-1}}\sqrt{1+r'(z)^2}=\int_\epsilon^{z_0}\dd z\;\mathcal L(r,r';z),
\end{equation}
where
\begin{equation}
    \mathcal L(r,r';z)=\mathrm{Vol}(S^{d-2})\frac{r(z)^{d-2}}{h(z)^{d-1}}\sqrt{1+r'(z)^2}.
\label{eq:mathcalL}
\end{equation}
Here $\epsilon$ is a UV cutoff and $z_0$ is the turning point of the minimal surface. The holographic setup is depicted in Figure \ref{fig:RTsurf}.

Since the inequalities derived from SSA in field theory contain derivatives of $S$, it will be convenient to alternatively parametrize the extremal surface by a profile $z(r)$.  In this case, the entropy is given by
\begin{equation}
    S(R) = \int_0^{R_c}\dd r\,\tilde\lgrn(z,\dot{z};r),
\end{equation}
where
\begin{equation}
    \tilde{\mathcal L}(z,\dot z;r)=\mathcal L(r,r';z)\,\fft{\dd z}{\dd r}=\mathrm{Vol}(S^{d-2})\;\frac{r^{d-2}}{h(z(r))^{d-1}}\sqrt{1+\dot z(r)^2}.
\end{equation}
Here, $\dot{z}=\dd z/\dd r$  and the surface is anchored such that $z(r=R_c)=\epsilon$. Although $S(R)$ is divergent on its own \cite{Witten:2018zxz}, after suitable subtraction of a reference entropy, $S(R)$ may be identified with the (relative) entropy of the dual field theory. In particular, it is important to emphasize that we have a well-defined holographic interpretation for all values of $R$, \emph{i.e.}, at all points along the RG flow.

Due to the minimality of the extremal surface, the bulk contributions to $S'$ disappear, leading to a Hamilton-Jacobi equation
\begin{align}
   S'(R) &= \dv{R_c}{R} \tilde\lgrn \bigg|_{R_c} + \int_0^{R_c}\dd r\left(\pdv{\tilde\lgrn}{z}\pdv{z}{R} + \pdv{\tilde\lgrn}{\dot{z}}\pdv{\dot{z}}{R}\right) \nonumber \\
   &= \dv{R_c}{R} \tilde\lgrn \bigg|_{R_c} + \pdv{\tilde\lgrn}{\dot{z}}\pdv{z}{R} \bigg|_0^{R_c},
\end{align}
where the equation of motion was used to reach the second line. At the cutoff, we have $z(R_c) = \epsilon$. Differentiating this condition gives
\begin{equation}
    \dot{z}\dv{R_c}{R}\bigg|_{R_c} + \pdv{z}{R}\bigg|_{R_c} = 0,
\end{equation}
and so
\begin{align}
    S'(R) &= \dv{R_c}{R}\left(\tilde\lgrn - \dot{z}\pdv{\tilde\lgrn}{\dot{z}}\right)\bigg|_{R_c} - \pdv{\tilde\lgrn}{\dot{z}}\pdv{z}{R}\bigg|_{0} \nonumber \\
    &=\mathrm{Vol}(S^{d-2})\qty(\dv{R_c}{R}\frac{r^{d-2}}{h^{d-1}\sqrt{1+\dot{z}^2}}\bigg|_{R_c} - \frac{r^{d-2}\dot{z}}{h^{d-1}\sqrt{1+\dot{z}^2}}\pdv{z}{R}\bigg|_{0}) \nonumber \\
    &= -\mathrm{Vol}(S^{d-2})\frac{r'r^{d-2}}{h^{d-1}\sqrt{1+r'^2}}\pdv{r}{R}\bigg|_{z=\epsilon},\label{S'}
\end{align}
where in the second line, the second term evaluates to zero.  In the final line, we have re-expressed $S'(R)$ in terms of the profile function $r(z)$.  Here we are considering a family of surfaces parametrized by $R$, and both $r'\equiv\partial r/\partial z$ and $\partial r/\partial R$ show up in this expression.  It is important to note that $S'(R)$ is a local function that depends only on the fields and their derivatives evaluated at the cutoff $z=\epsilon$. Note also that \eqref{S'} matches the Hamilton-Jacobi equation obtained in \cite{Daguerre:2022uxt}, up to a change of coordinates.\footnote{However, it is important to note that Ref. \cite{Daguerre:2022uxt} uses a different anchoring prescription, which is equivalent to $z(R)=\epsilon$, rather than $z(R_c)=\epsilon$.}
Similar results were obtained using the Jacobi equation in \cite{Speranza:2019hkr,Engelhardt:2019hmr}.

\subsection{Equations of motion for the RT surface}

Although $S'(R)$ can be evaluated locally at the boundary, we still need to understand the behavior of the surface at the boundary.  To do so, we turn to the equation of motion for the minimal surface
\begin{equation} \label{EOM_d}
    r''(z) = \left(\frac{d-2}{r(z)} + (d-1)\frac{h'(z)r'(z)}{h(z)}\right)\left(1+r'(z)^2\right).
\end{equation}
For $d=2$ (\textit{i.e.},~AdS$_3$), this admits a first integral, $r'(z)=h(z)/\sqrt{h(z_0)^2-h(z)^2}$.  However, no general closed-form solution is known for $d\ge3$, except for the semicircle solution, $r(z)=\sqrt{z_0^2-z^2}$, in the conformal case where $h(z)=z$.  Nevertheless, one may construct series expansions near the boundary $z=0$ and turning point $z=z_0$. The surface profile has boundary conditions $r(0)=R$ and $r(z_0)=0$.

Without bulk matter, the canonical way to examine the behavior of the metric in asymptotically AdS spacetimes is via the Fefferman-Graham expansion \cite{Fefferman:1985}; however, in the presence of a relevant deformation by an operator of conformal dimension $\Delta_-=d-\Delta_+$, we will generically have backreaction that leads to a metric boundary expansion of the form%
\footnote{Here and throughout this manuscript, we use the little-$o$ notation where $o(\cdot)$ denotes growth which is strictly subleading compared to the argument in parentheses. This is in contrast to $\mathcal O(\cdot)$, which denotes behavior that may be of the same order or subleading with respect to the argument in parentheses.}
\begin{equation}
    h(z)=z+c z^{1+2\Delta_+}+o\qty(z^{1+2\Delta_+}).\label{eq:metric-exp}
\end{equation}
Here $\Delta_+$ is the conformal dimension of the perturbing coupling, and for generic $\Delta_+$ the boundary expansion will be non-analytic. This is summarized explicitly for a scalar deformation in Appendix \ref{app:DWflows}. Note that we must have $\Delta_+>0$ to get asymptotic AdS, or, equivalently, $\Delta_-<d$. This is simply the requirement that the deformation is relevant. Note also that the NEC requires $c\ge 0$.

In the absence of such a perturbation, \emph{i.e.}, in the case of pure AdS, the minimal surface is just the semicircle
\begin{equation}
    r_0(z)=\sqrt{R^2-z^2}.
\end{equation}
More generally, the geometry near the boundary takes the form
\begin{equation}
    r(z)=r_0(z)+r_d(R)z^d+\zeta_1(R)z^{2+2\Delta_+}+o(z^{2+2\Delta_+})+\mathcal{O}(z^{d+2\Delta_+}),\label{r(z) form}
\end{equation}
where
\begin{equation}
    \zeta_1(R)=\frac{(d-1)\Delta_+}{(d-2-2\Delta_+)(\Delta_++1)}\frac{c}{R}.\label{eq:zeta}
\end{equation}
Here, $r_d(R)$ is an integration constant that corresponds to global information about the turning point, the $z^{2+2\Delta_+}$ term is the leading correction due to the backreaction of the metric, and the $\mathcal{O}(z^{d+2\Delta_+})$ terms include the cross terms between the $r_d(R)$ term and the metric corrections. Note that in the Fefferman-Graham expansion, $r_d$ precisely corresponds to $(r^{(d)}(0)-r^{(d)}_0(0))/d!$. When the entangling region is small, $r_d$ may be found explicitly (see Appendix \ref{app:rd}).

Despite the possible non-analyticity near the boundary, the metric expansion around a bulk point is simply a Taylor series, and hence the expansion of $z(r)$ near the $r=0$ turning point is more straightforward:
\begin{equation} \label{r_expan_d}
    z(r)=z_0 -\frac{h'(z_0)}{2 h(z_0)}r^2 + \left(\frac{(d-1)h'(z_0) h''(z_0)}{8(d+1)h(z_0)^2}-\frac{h'(z_0)^3}{8h(z_0)^3}\right)r^4 + \mathcal{O}(r^6).
\end{equation}
This will be useful to us later.

\subsection{An approximate constant of motion}

In pure AdS, corresponding to unbroken conformal invariance, the RT surface is just a semicircle.  Moreover, there is a constant of motion for the surface associated with scale invariance: $z\to \lambda z,\,\, r\to \lambda r$. By Noether's theorem, this constant is%
\footnote{Here we have removed the constant volume factor, $\mathrm{Vol}(S^{d-2})$, from $\mathcal C$ and will continue to do so below.}
\begin{equation}
    \mathcal{C} = \left(\lgrn - \frac{\partial \lgrn}{\partial r'}r'\right)z + \frac{\partial \lgrn}{\partial r'}r = \frac{r^{d-2}(z+rr')}{z^{d-1}\sqrt{1+(r')^2}} = 0,
\end{equation}
where $\mathcal L$, given in (\ref{eq:mathcalL}), corresponds to the $r(z)$ parametrization of the surface.  For a flow in which conformal symmetry is broken, the above quantity will instead vary along the RT surface as a function of $z$. However, a generalized form of $\mathcal{C}(z)$, though non-constant, will act as a measure of conformal symmetry breaking and produce useful constraints on the behavior of the surface. 

At any point $z=z^*$, we may consider the linear approximation for $h(z)$ given by
\begin{equation}
     \bar h(z)=h(z^*)+h'(z^*)(z-z^*),
\end{equation}
Since any metric with linear $h(z)$ describes AdS, $\bar h(z)$ is effectively an AdS approximation to the flowing theory at $z^*$. We may then consider the conserved quantity derived from the scale symmetry of this associated AdS metric. In the UV and IR limits, the linear approximation will become exact, and the resulting quantity will be conserved in the flowing theory as well. In the intermediate regime, it will be non-constant.

\begin{figure}[t]
\centering
\tikzset{every picture/.style={line width=0.75pt}} 

\begin{tikzpicture}[x=0.75pt,y=0.75pt,yscale=-0.75,xscale=0.75]

\draw    (11,409.6) -- (558,409.6) ;
\draw [shift={(560,409.6)}, rotate = 180] [color={rgb, 255:red, 0; green, 0; blue, 0 }  ][line width=0.75]    (10.93,-3.29) .. controls (6.95,-1.4) and (3.31,-0.3) .. (0,0) .. controls (3.31,0.3) and (6.95,1.4) .. (10.93,3.29)   ;
\draw    (10.97,409.6) -- (11,19.6) ;
\draw [shift={(11,17.6)}, rotate = 90.01] [color={rgb, 255:red, 0; green, 0; blue, 0 }  ][line width=0.75]    (10.93,-3.29) .. controls (6.95,-1.4) and (3.31,-0.3) .. (0,0) .. controls (3.31,0.3) and (6.95,1.4) .. (10.93,3.29)   ;
\draw [color={rgb, 255:red, 74; green, 144; blue, 226 }  ,draw opacity=1 ]   (11,409.6) .. controls (424,289.6) and (351,343.6) .. (539,27.6) ;
\draw [color={rgb, 255:red, 208; green, 2; blue, 27 }  ,draw opacity=1 ] [dash pattern={on 3.75pt off 3pt on 7.5pt off 1.5pt}]  (539,27.6) -- (313,408.6) ;
\draw  [dash pattern={on 0.84pt off 2.51pt}]  (445,188.6) -- (444,408.6) ;

\draw (565,399.4) node [anchor=north west][inner sep=0.75pt]    {$z$};
\draw (7,-0.6) node [anchor=north west][inner sep=0.75pt]    {$h$};
\draw (306,413.4) node [anchor=north west][inner sep=0.75pt]    {$z_{c}$};
\draw (437,414.4) node [anchor=north west][inner sep=0.75pt]    {$z^{*}$};
\draw (6,414.4) node [anchor=north west][inner sep=0.75pt]    {$0$};
\end{tikzpicture}
\caption{The warp factor $h$ is shown (schematically) in blue, and its linearization $\bar h$ around the point $z^*$ is shown in red. $z_c$ naturally arises as the point where $\bar h$ intersects the $z$-axis.}
\label{fig:zc}
\end{figure}
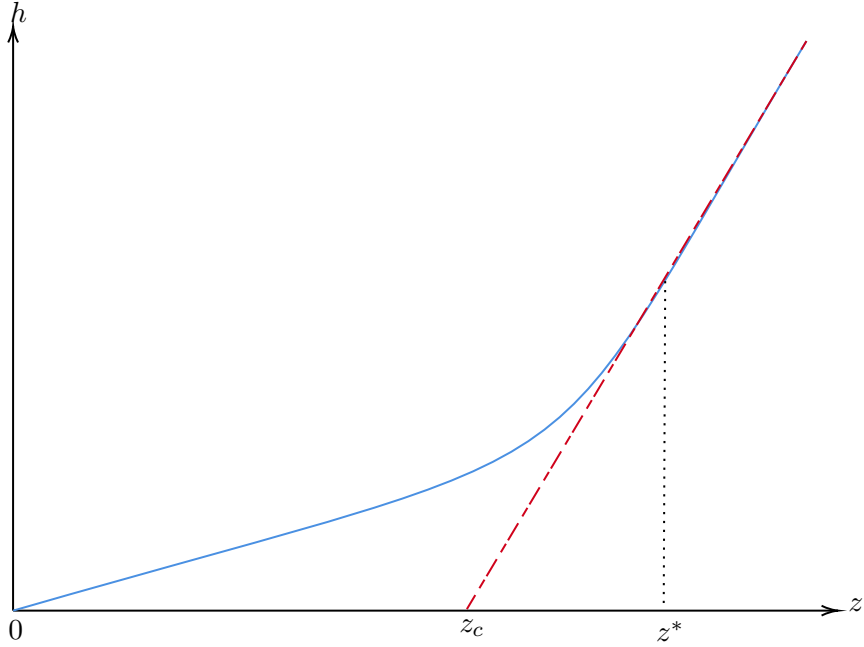

Since the linear approximation does not in general satisfy $\bar{h}(0)=0$, we should consider dilatations centered around
\begin{equation}
     z_c = z^*-\frac{h(z^*)}{h'(z^*)},
\end{equation}
where $z_c$ satisfies $\bar h(z_c)=0$. This is depicted schematically in Figure \ref{fig:zc}.  Infinitesimal dilatations with scale factor $(1+\delta\lambda)$ then take the form
\begin{align}\label{eqn:dilat}
    &r\;\to\;r + \delta\lambda\,r, \nonumber\\
    &z\;\to\;z + \delta\lambda\,(z-z_c) = z + \delta\lambda\left(z-z^* + \frac{h(z^*)}{h'(z^*)}\right).
\end{align}
Following Noether's theorem, we define
\begin{equation}
    \mathcal{C}(z) = \left(\lgrn - \frac{\partial \lgrn}{\partial r'}r'\right)\frac{h(z)}{h'(z)} + \frac{\partial \lgrn}{\partial r'}r = r^{d-2}\frac{h+h'rr'}{h'h^{d-1}\sqrt{1+r'^2}},
\end{equation}
where for each $z$ we take the generator to be that of the dilatation defined by the linear approximation at that point (by setting $z=z^*$ in \eqref{eqn:dilat}). Since $h''(z)$ describes the deviation of the flowing theory from pure AdS, it should also control the variation of $\mathcal{C}(z)$. Indeed, imposing the equations of motion \eqref{EOM_d} yields
\begin{equation} \label{qconst ineq}
    \dv{\mathcal{C}}{z} = -\frac{r^{d-2}h''}{(h')^2h^{d-2}\sqrt{1+r'^2}} \leq 0,
\end{equation}
where the inequality results from the NEC \eqref{eq:NEC}. Therefore, $\mathcal{C}(z)$ is non-increasing everywhere and zero if and only if the NEC is saturated.
Near the turning point, the expansion \eqref{r_expan_d} gives
\begin{equation}
    \mathcal{C}(z(r)) = \frac{h''(z_0)}{(d+1)h(z_0)^{d}}\,r^{d+1} + \mathcal{O}\qty(r^{d+3}),
\end{equation}
so $\mathcal{C}(z_0) = 0$. Thus $\mathcal{C}$ is always non-negative, and in particular
\begin{equation}
    h+h'rr' \geq 0.\label{eq:Cineq}
\end{equation}
This inequality will be used in the proof below.

\subsection{Entropic inequalities}

Field theoretically, the authors of \cite{Casini:2017vbe} used strong subadditivity to show that a suitably subtracted EE, $\Delta S_\text{EE}(R)$, satisfies the inequality \eqref{Eq:MainSubAd}, namely
\begin{equation}
    R\Delta S''(R)-(d-3)\Delta S'(R)\le0,
\end{equation}
which may be rewritten as
\begin{equation}
    R^{d-2}\dv{}{R} \left(\frac{\Delta S'}{R^{d-3}}\right)\le0.
\end{equation}
To reproduce this result holographically, we must determine the boundary behavior of the minimal surfaces and utilize the expression \eqref{S'} for $S'(R)$. We now consider the $z$-evolution of the infinitesimal separation between two minimal surfaces. The differential equation describing this separation is sufficient to derive the desired monotonicity properties.

From \eqref{S'} we may write
\begin{align}
\dv{}{R} \left(\frac{\Delta S'}{R^{d-3}}\right) &= -\frac{\mathrm{Vol}(S^{d-2})}{h^{d-1}}\dv{}{R}\left(rr'\frac{r^{d-3}}{R^{d-3}\sqrt{1+r'^2}}\dv{r}{R}\right)\bigg|_{z=\epsilon}\nonumber\\
&\qquad+ \frac{\mathrm{Vol}(S^{d-2})}{z^{d-1}}\dv{}{R}\left(r_0 r_0'\frac{r_0^{d-3}}{R^{d-3}\sqrt{1+r_0'^2}}\dv{r_0}{R}\right)\bigg|_{z=\epsilon}.\label{eq:cfunc}
\end{align}
The first term in this difference originates from the entropy of the flowing theory, while the second term is the subtraction of the UV CFT contribution. Here, a factor of $rr'$ has been separated out inside the first set of parentheses, and the leftover part in the parentheses is just 1 to leading order. A similar case holds for the CFT portion. Hence, one may expect that $rr'$ will give the major contribution to the overall expression after subtraction. In fact, we will demonstrate that
\begin{equation}
    \dv{}{R} \left(\frac{\Delta S'}{R^{d-3}}\right)\sim -\frac{1}{h^{d-1}}\dv{}{R}\qty(rr')\bigg|_{z=\epsilon},
\end{equation}
to leading order in $\epsilon$.

At this point, it is interesting to note that we may rewrite the above derivative as
\begin{equation}
    \dv{}{R}\qty(rr')=r'\dv{r}{R}+r\dv{r'}{R}=\dv{}{z}\qty(r\dv{r}{R}).
\end{equation}
This then motivates us to consider the quantity
\begin{equation}
    v(z):=\frac{r}{R}\dv{r}{R}.
\end{equation}
Intuitively, $\dd{r}/\dd{R}$ measures the $r$-separation between neighboring surfaces, and the $r/R$ factor keeps $v(z)$ finite at the turning point. In pure AdS, $v(z)$ is constant since, in this case, it amounts to a measurement of the perpendicular separation between surfaces with respect to a flat metric in the $r$-$z$ plane. Thus, one may interpret $v(z)$ as linearizing the deviation of surfaces. Now, we would like to constrain the sign of $v'(z)$, which leads us to the following Lemma.

\medskip
\noindent\textbf{Lemma}: The NEC implies that the separation $v(z)$ never decreases.
$$h''(z)\geq0 \quad\implies\quad  v'(z)\geq 0.$$
\textbf{Proof}:
We proceed by demonstrating two results:
\begin{itemize}
    \item $v(z)$ has no local minima.
    \item Smoothness at the turning point requires $v'(z) > 0$ for physically reasonable flows, so $v'(z) \geq 0 $ for all $z\in[0,z_0]$.
\end{itemize}
If we take the $R$ derivative of the minimal surface equation, \eqref{EOM_d}, and rewrite in terms of $v(z)$, we get a linear second-order ODE
\begin{equation}
\label{v_diffeq_d}
    v'' - (d-1)\left(\frac{2 r'}{r} + \frac{h'}{ h}\left(1 + 3 r'^2\right)\right)v' +2(d-1)\frac{r'^2}{h r^2}\left(h + h'rr'\right)v = 0.
\end{equation} 

Notice that the coefficient of $v$ in \eqref{v_diffeq_d} is non-negative by \eqref{eq:Cineq}. By the monotonicity property of $\mathcal{C}(z)$, if $h+h'rr' = 0$ at any value of $z<z_0$ it must be zero everywhere and correspond to the case of pure AdS. For nontrivial $h(z)$, the coefficient of the third term of \eqref{v_diffeq_d} is therefore strictly positive for $z<z_0$. Separate minimal surfaces in the family cannot intersect,\footnote{See Figure 10 and Theorem 17 (b)-(d) in \cite{Wall:2012uf}. The theorem relates to a more general situation involving maximin surfaces, but the mentioned steps of the proof apply to our case as well.} so $v(z)$ should be positive. Therefore, any critical point $z_c \in (0,z_0)$ satisfying $v'(z_c)=0$ has $v''(z_c)<0$. Local minima are thus forbidden.

Now, consider the behavior at the turning point. One may na\"ively expect that smoothness requires $v'(z_0) = 0$, but the singular behavior of the $z$ parametrization at the turning point leads to a different result. We then substitute the $r=0$ expansion \eqref{r_expan_d} into \eqref{v_diffeq_d}, and use $r'(z(r)) = 1/\dot{z}(r)$. Setting each term of the $r$ expansion of \eqref{v_diffeq_d} to zero constrains the derivatives of $v$ at $z=z_0$. The expansion of $v(z)$ is then
\begin{equation}
    \frac{v(z)}{v(z_0)} = 1 + \frac{2 h''(z_0)}{(d+1) h'(z_0)}(z-z_0) + \mathcal{O}\left((z-z_0)^2\right).
\end{equation}

Recall that the NEC implies $h''(z_0)\geq 0$. For domain wall flows between two fixed points, $h''(z)$ is expected to approach zero only as a limit in the IR. In this case, we have the strict inequality $v'(z) > 0$ in the neighborhood of $z=z_0$. Since local minima cannot exist, $v'(z) \geq 0$ for all $z<z_0$. $\square$

Given the above Lemma, the quick proof is that
\begin{equation}
    \dv{}{R} \left(\frac{\Delta S'}{R^{d-3}}\right)\sim-\frac{\mathrm{Vol}(S^{d-2})}{h^{d-1}}v'\bigg|_{z=\epsilon}\le 0
\end{equation}
to leading order. Of course, we must ensure that there are no other divergences that ruin this statement, which we check now.

We may evaluate \eqref{eq:cfunc} given our metric \eqref{eq:metric} and the boundary expansion \eqref{r(z) form} to write
\begin{align}
    \dv{}{R} \left(\frac{\Delta S'}{R^{d-3}}\right)&=-\mathrm{Vol}(S^{d-2})\Bigg[\qty(d\dv{}{R}\qty(Rr_d(R))+\mathcal{O}(\epsilon^{2\Delta_+}))\nonumber\\
    &+\qty(\frac{2d^2(1+\Delta_+)-4d(2+\Delta_+)+4(2+\Delta_++2\Delta_+^2)}{(d-2-2\Delta_+)(2+2\Delta_+)}\frac{(d-1)c}{R^3}\epsilon^{4-d+2\Delta_+}+o(\epsilon^{4-d+2\Delta_+}))\Bigg].\label{eq:entropyderiv}
\end{align}
Now, several important points are in order. First, there is a subtle interplay between factors of $R$ and $z$.  At this level, $r_d(R)$ is a generic function of $R$ and so we can be assured that all corrections are at least $\mathcal{O}(\epsilon^{2\Delta_+})$; however, the $z^{2+2\Delta_+}$ term in $r(z)$ has a coefficient which is a non-generic function of $R$, so extra caution is required. Indeed, if instead of the $R$ dependence following from Equations  \eqref{r(z) form} and \eqref{eq:zeta}, namely $\zeta_1(R)\propto R^{-1}$, we had any other $R$ dependence, then we would have to contend with additional, divergent 
 $\epsilon^{2-d+2\Delta_+}$ terms proportional to $\dv{}{R}\qty(R\zeta_1(R))$. Thus, to ensure that our power counting is indeed correct, one must additionally check that the subleading terms also have a $R^{-1}$ dependence. Fortunately, this is indeed the case. If one writes the more general expression
\begin{equation*}
    h(z)=z+\sum_{n=1}^\infty c_n z^{1+2n\Delta_+}+\sum_{n=2}^\infty \ell_n z^{1+2n\Delta_+}\log z,
\end{equation*}
\begin{equation}
    r(z)=r_0(z)+\qty(\frac{r_d(R)}{R}z^d+\mathcal{O}(z^{d+2\Delta_+}))+\qty(\sum_{n=1}^\infty \zeta_n(R) z^{2+2n\Delta_+}+\sum_{n=2}^\infty \xi_n(R) z^{2+2n\Delta_+}\log z)\Bigg],
\end{equation}
and expands the equations of motion in $z$, then we get
\begin{equation}
    0=(d-3)\sigma+(d-1)z\sigma'+(d-1)R\Sigma'+(d-1)R\sigma'\Sigma'-zR\Sigma''-R\sigma\Sigma''+\mathcal{O}(z^{1+2n_\mathrm{max}\Delta_+}),
\label{eq:algeqn}
\end{equation}
where
\begin{align}
    \sigma&=\sum_{n=1}^{n_\mathrm{max}} c_n z^{1+2n\Delta_+}+\sum_{n=2}^{n_\mathrm{max}} \ell_n z^{1+2n\Delta_+}\log z,\nn\\
    \Sigma&=\sum_{n=1}^{n_\mathrm{max}} \zeta_n(R) z^{2+2n\Delta_+}+\sum_{n=2}^{n_\mathrm{max}} \xi_n(R) z^{2+2n\Delta_+}\log z,
\end{align}
and $n_\mathrm{max}=\lfloor\tfrac{1}{\Delta_+}\rfloor$.%
\footnote{The $R$ dependence becomes difficult to manage once we have cross terms with the subleading terms of $r_0=R-z^2/2R+\cdots$ or with $r_d(R)z^d$. Since we are interested in $d\ge 2$, the $r_d$ term is, at worst, the same order as $z^2/2R$. Hence, we have control until $2n\Delta_+=2$, at which point the first subleading term $z^2/2R$ of $r_0$ is the same order. Thus, $n_\mathrm{max}=\lfloor\tfrac{1}{\Delta_+}\rfloor$.}
Thus, we see from (\ref{eq:algeqn}) that $\zeta_n(R)$ and $\xi_n(R)$ always enter with a factor of exactly $R$. Hence, although this algebraic equation is difficult to solve generally, it is clear that we will always get
\begin{equation}
    \zeta_n,\,\xi_n\propto \frac{1}{R},\qquad n\le n_{\mathrm{max}}.
\end{equation}
Thus, our order counting above is sufficient (up to $\epsilon^{4-d+2\Delta_+}$), and this is then a second source of $\epsilon^{4-d+2\Delta_+}$ terms.


Now, this naturally leads to an issue for sufficiently large $d$: If $d-4>2\Delta_+$, then our expression is divergent. This suggests that there is a lower bound on $\Delta_+$ for the expression to make sense
\begin{equation}
    \Delta_+\ge \frac{d-4}{2}.
\end{equation}

To prove monotonicity, we want to connect the quantity \eqref{eq:entropyderiv} to $v(z)$. It is straightforward to expand around $z=0$ as
\begin{equation}
    v(z)=1+\qty(\frac{1}{R}\dv{}{R}\qty(R r_d)z^d+o\qty(z^d))+\qty(\frac{2(d-1)\Delta_+}{(d-2-2\Delta_+)(2+2\Delta_+)}\frac{2c}{R^4}z^{4+2\Delta_+}+o\qty(z^{4+2\Delta_+})).
\end{equation}
Note that the bound $2\Delta_+\ge d-4$ automatically implies that $z^{4+2\Delta_+}$ is, at worst, the same order as $z^d$ (note also that the assumption that $\Delta_+>0$ means that $4+2\Delta_+ > d$ for $d\le 4$, so the possibility of saturating this bound is only an issue for $d>4$). Thus, from the Lemma, we conclude that \eqref{eq:entropyderiv} is non-positive, and this suffices to prove the monotonicity of \eqref{eq:cfunc}. In particular, this provides a holographic proof of the $c$-, $F$-, and $a$-theorems, analogous to the field-theoretic version \cite{Casini:2017vbe}. For $d\ge 5$, the function \eqref{eq:cCTT} no longer corresponds to a central charge at the fixed points, but it does still give us some sort of monotonic quantity.


\section{Flows across dimensions}\label{sec:acrossDims}
Given the proof of monotonicity in the preceding section for flows within the same dimension, it is natural to attempt to generalize this to the case of flows across dimensions. Hence, we choose a metric
\begin{equation}
    \dd s^2=\frac{1}{h(z)^2}\qty(-\dd t^2+\dd z^2+\dd r^2+r^2\dd\Omega_{d-2}^2)+\frac{1}{j(z)^2}\dd s^2_{M_{D-d}},\label{eq:metFlowAcDims}
\end{equation}
with asymptotics
\begin{align}
    h(z)\overset{z\to 0}{\sim}z,&\qquad h(z)\overset{z\to\infty}{\sim}\frac{z}{L_\mathrm{IR}},\nonumber\\
    j(z)\overset{z\to 0}{\sim}z,&\qquad j(z)\overset{z\to\infty}{\sim}j_\mathrm{IR},
\end{align}
where $j_\mathrm{IR}$ is a constant. This metric then interpolates between a UV $\mathrm{AdS}_{D+1}$ and an IR $\mathrm{AdS}_{d+1}\times M_{D-d}$, which is holographically dual to a flow from a $\mathrm{CFT}_D$ to a $\mathrm{CFT}_d$, compactified on $M_{D-d}$ \cite{GonzalezLezcano:2022mcd}. Note that the special case $D=d$ corresponds to the metric \eqref{eq:metric}. The metric \eqref{eq:metFlowAcDims} has Ricci tensor components
\begin{align}
    \hat R^\mu{}_\nu&=\qty(hh''-d(h')^2-(D-d)hh'\frac{j'}{j})\delta^\mu_\nu,\nonumber\\
    \hat R^i{}_j&=j^2R^i{}_j+\frac{h}{j^2}\qty(-(D-d-1)h(j')^2+j\qty(hj''-(d-1)h'j'))\delta^i_j,\nonumber\\
    \hat R^z{}_z&=d\qty(hh''-(h')^2)+(D-d)\frac{h}{j^2}\qty(-2h(j')^2+j\qty(h'j'+hj'')).
\end{align}
We assume that the internal space $M_{D-d}$ is Einstein with metric $g_{ij}$ and curvature
\begin{equation}
    R_{ij}=(D-d-1)\frac{\kappa}{\ell^2}g_{ij},
\end{equation}
where $\kappa = 1,$ 0, or $-1$ for positive, flat, or negative curvature, respectively, and $\ell$ characterizes the scale of the compact dimensions. Since the $D$-dimensional isometry is broken by the flow, we end up with two independent NEC inequalities
\begin{align}
    \text{NEC1:}\qquad&\tilde h''(z) \geq\frac{(D-1)(D-d)}{(d-1)^2}\tilde h(z)\qty(\frac{j'(z)}{j(z)})^2 \geq 0,\nonumber\\
    \text{NEC2:}\qquad&d\qty(\frac{h'}{h})^2+\frac{(D-2d+1)h'j'-jh''}{hj}+\frac{j''j-(D-d+1)(j')^2}{j^2}+(D-d-1)\frac{\kappa}{\ell^2}\frac{j^2}{h^2}\ge0,
\end{align}
where we have defined the effective warp factor
\begin{equation}
    \tilde h(z):=h(z) j(z)^\frac{D-d}{d-1}.
\end{equation}
While it is possible to violate the NEC, especially in the presence of compact dimensions (see \emph{e.g.} \cite{Kontou:2020bta} for a discussion of energy conditions and their violation), we will focus on the case where the bulk NEC is satisfied.

We will consider entanglement regions that wrap $M_{D-d}$ and are spherical in the remaining $d-1$ dimensions, with radius $R$. In particular, the entropy functional
\begin{equation}
    S(R) = \mathrm{Vol}(S^{d-2})\mathrm{Vol}(M_{D-d})\int_\epsilon^{z_0}\dd z\;\frac{r(z)^{d-2}}{\tilde h(z)^{d-1}}\sqrt{1+r'(z)^2},\label{eq:acrossdimsS}
\end{equation}
and NEC1 appear almost identical to the case of flows within the same dimension. Hence, most of the calculations proceed exactly as before with the replacement $h\to\tilde h$. In particular, we still have a monotonically decreasing function
\begin{equation}
    \mathcal{C}(z) = r^{d-2}\frac{\tilde h+\tilde h'rr'}{\tilde h'\tilde h^{d-1}\sqrt{1+r'^2}},
\end{equation}
and the Lemma straightforwardly generalizes to flows across dimensions, \emph{i.e.}, $v'(z)\ge0$. Likewise, we find
\begin{align}
\dv{}{R} \left(\frac{\Delta S'}{R^{d-3}}\right) &= -\frac{\mathrm{Vol}(S^{d-2})\mathrm{Vol}(M_{D-d})}{\tilde h^{d-1}}\dv{}{R}\left(rr'\frac{r^{d-3}}{R^{d-3}\sqrt{1+r'^2}}\dv{r}{R}\right)\bigg|_{z=\epsilon}\nonumber\\
&\qquad + \frac{\mathrm{Vol}(S^{d-2})\mathrm{Vol}(M_{D-d})}{\tilde h_0(z)^{d-1}}\dv{}{R}\left(r_0 r_0'\frac{r_0^{d-3}}{R^{d-3}\sqrt{1+r_0'^2}}\dv{r_0}{R}\right)\bigg|_{z=\epsilon},
\end{align}
where $\tilde h_0=z^{(D-1)/(d-1)}$ corresponds to the non-flowing geometry with $h_0=j_0=z$.

The main place where the distinction from flows within the same dimension becomes clear is when considering expansions around the boundary since $\tilde h$ has asymptotics
\begin{equation}
    \tilde h(z)\overset{z\to 0}{\sim}z^\frac{D-1}{d-1}.
\end{equation}
Not only that, but the internal space $M_{D-d}$ itself may contribute to the curvature when $\kappa$ is non-vanishing. This leads to corrections to the metric, even in the absence of any relevant perturbations. Solving the vacuum AdS Einstein equations
\begin{equation}
    \hat R_{MN}-\frac{1}{2}\hat g_{MN}+\frac{D(D-1)}{2}=0,
\end{equation}
we may find the effective warp factor as a series expansion
\begin{equation}
    \tilde h_0(z)=z^{\frac{D-1}{d-1}}\qty(1+\frac{(D-d)(D-d-1)}{3(D-2)(d-1)}\frac{\kappa}{\ell^2}z^2+\mathcal{O}\qty(\frac{\kappa}{\ell^2} z^4)),\label{eq:curvmetric}
\end{equation}
although the full solution is not known in general. As we will see, this leads to some important differences. 

\subsection{Flows to AdS$_3$}
As a warm-up, we use our new formalism to demonstrate the monotonicity of flows from $\mathrm{AdS}_{D+1}$ to $\mathrm{AdS}_3$, which was proved originally in \cite{GonzalezLezcano:2022mcd}. We start by finding the unperturbed minimal surface profile $r_0(z)$. Na\"ively, we might like to simply solve the equations of motion perturbatively, order by order. Unfortunately, this gives the unsatisfactory solution $r_0=R$, which is not a closed surface. Fortunately, we may use the integral of motion to find $r_0(z)$.

\begin{figure}\label{fig:r0}
    \centering
    \includegraphics{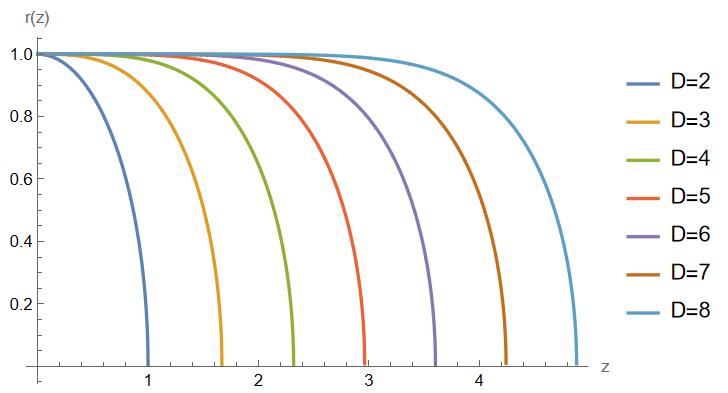}
    \caption{The profile $r_0(z)$ for several values of $D$, for the case $\kappa=0$. Notice that increasing $D$ elongates the profile. We have set $R=1$ for simplicity of visualization.}
\end{figure}

The entropy functional \eqref{eq:acrossdimsS} has no explicit $r(z)$ factors for $d=2$, which implies the existence of a constant of motion given by
\begin{equation}
    \frac{r'}{\tilde h\sqrt{1+(r')^2}}=\text{const.}
\end{equation}
After rearranging and imposing the condition that $\lim_{z\to z_0}r'(z)=-\infty$, we find that
\begin{equation}
    r'(z)=-\frac{\tilde h(z)/\tilde h(z_0)}{\sqrt{1-(\tilde h(z)/\tilde h(z_0))^2}},
\end{equation}
which, for the torus case, \emph{i.e.}, $\kappa=0$, becomes simply
\begin{equation}
    r_0(z)=\frac{\sqrt{\pi}\Gamma\qty(\frac{D}{2D-2})}{\Gamma\qty(\frac{1}{2D-2})}z_0-\frac{z_0}{D}\qty(\frac{z}{z_0})^D{}_2 F_1\qty(\frac{1}{2},\frac{D}{2D-2};1+\frac{D}{2D-2};\qty(\frac{z}{z_0})^{2(D-1)}).
\end{equation}
Here the turning point $z_0$ is implicitly a function of $R$ given by the requirement that $r_0(0)=R$, which may be solved to give
\begin{equation}
    z_0=\frac{\Gamma\qty(\frac{1}{2D-2})}{\Gamma\qty(\frac{D}{2D-2})}\frac{R}{\sqrt{\pi}},
\end{equation}
and so we get the behavior
\begin{equation}
    r_0(z)= R-\frac{z_0}{D}\qty(\frac{z}{z_0})^D+\mathcal{O}(z^{2D}).
\end{equation}
The profile $r_0$ is shown in Figure \ref{fig:r0}.
Note that for $D=2$, this reduces precisely to the semicircle $r_0=\sqrt{R^2-z^2}$ with turning point $z_0=R$. 

When $\kappa\ne 0$, we must use the more general metric \eqref{eq:curvmetric}. This leads to a more general expression for $r_0(z)$:
\begin{align}
    r_0(z)&=R-\frac{\tilde h(z_0)}{D}\qty(\frac{z}{\tilde h(z_0)})^D{}_2 F_1\qty(\frac{1}{2},\frac{D}{2D-2};\frac{3D-2}{2D-2};\qty(\frac{z}{\tilde h(z_0)})^{2(D-1)})\nonumber\\
    &\quad-\frac{(D-1)(D-3)}{3(D+2)}\frac{\kappa}{\ell^2}\frac{z^{D+2}}{\tilde h(z_0)^{D-1}}{}_2 F_1\qty(\frac{1}{2},\frac{D+2}{2D-2};\frac{3D}{2D-1},-\qty(\frac{z}{\tilde h(z_0)})^{2(D-1)})\nonumber\\
    &\quad+\mathcal{O}\qty(z^{D+4})\nonumber\\
    &= R-\frac{z_0}{D}\qty(\frac{z}{z_0})^D+\mathcal{O}(z^{D+2}).
\end{align}
The main distinction is that in the case $\kappa\ne 0$, there is no longer a simple expression for the turning point of $r_0$. Instead, we will leave it implicit as a function $z_0(R)$.

Hence, we write
\begin{equation}
    r(z)=r_0(z)+r_D(R) z^D+\mathcal{O}(z^{D+2\Delta_+}).
\end{equation}
Importantly, both the metric backreaction and internal curvature pieces are subleading compared to $r_D$. Hence, we may compute the expansion of $v$ to give
\begin{equation}
    v(z)=1+\qty(-\frac{1}{DR}\dv{}{R}\qty(\frac{R}{z_0^{D-1}})+\frac{1}{R}\dv{}{R}\qty(R\,r_D))z^D+\mathcal{O}(z^{D+2\Delta_+}).
\end{equation}
For the torus case, this first term is simply
\begin{equation}
    -\frac{1}{DR}\dv{}{R}\qty(\frac{R}{z_0^{D-1}})\overset{\kappa\to 0}{=}\frac{D-2}{D}\frac{\pi^{(D-1)/2}}{R^{D}}\qty(\frac{\Gamma\qty(\frac{D}{2D-2})}{\Gamma\qty(\frac{1}{2D-2})})^{D-1}.
\end{equation}

We may then compute the \emph{unsubtracted} central charge derivative to find that
\begin{equation}
    \dv{}{R} \left(RS'\right) = -2D\mathrm{Vol}(M_{D-2})\qty(-\frac{1}{DR}\dv{}{R}\qty(\frac{R}{z_0^{D-1}})+\frac{1}{R}\dv{}{R}\qty(R\,r_D))+\mathcal{O}(\epsilon^{2\Delta_+}).
\end{equation}
Note that we have formally taken $\mathrm{Vol}(S^0)=2$, as it consists of the two endpoints of the minimal curve.  Hence, given that $v'(z)\ge0$ by the Lemma, we have the monotonicity of flows to $d=2$, namely to $\mathrm{AdS}_3$.

One may wonder why it is sufficient to use the unsubtracted central charge, as the entanglement entropy, $S(R)$, is UV divergent in $D$ dimensions.%
\footnote{It should be noted that the original proof of \cite{GonzalezLezcano:2022mcd} did so with the unsubtracted central charge as well.}
The key point is that because $D-2$ directions are compact, and the minimal surface wraps the entire compact dimension while terminating at two isolated points on the spatial boundary of AdS$_3$, the power-law divergences will be independent of the separation $R$.  In other words, the entanglement entropy for the UV CFT takes the form
\begin{equation}
    S_\mathrm{UV}=\mathrm{Vol}(M_{D-2})\qty[\frac{\mu_{D-2}}{\epsilon^{D-2}}+\frac{\mu_{D-4}}{\ell^2\epsilon^{D-4}}+\cdots+a_\mathrm{UV}\log\frac{R}{\epsilon}+\mathcal O(\epsilon^0)],
\end{equation}
where the coefficients $\mu_i$ are independent of $R$.  As a result, taking a derivative, $S'=\dd S/\dd R$, removes all UV divergences, and the extracted central charge is finite. This, of course, is specific to the case of flows to $d=2$.

\subsection{Flows to $\mathrm{AdS}_{d+1}$}
Now we consider the general case of flows with $D>d>2$. In this case, the equations of motion corresponding to minimizing the entropy functional \eqref{eq:acrossdimsS} are
\begin{equation}
    r''(z) = \left(\frac{d-2}{r(z)} + (d-1)\frac{\tilde h'(z)r'(z)}{\tilde h(z)}\right)\left(1+r'(z)^2\right).
\end{equation}
Unfortunately, no closed-form solution is known for $r_0$, even when $\kappa=0$. Instead, we must expand perturbatively around $z=0$, which gives an expansion
\begin{align}
    r_0(z)&=R-\frac{1}{2}\frac{d-2}{D-2}\frac{z^2}{R}\nonumber\\
    &-\frac{d-2}{6(D-2)^3(D-4)}\qty(\frac{3(d-2)(D^2-4D-2d+8)}{R^2}-4(D-d)(D-d-1)(D-2)\frac{\kappa}{\ell^2})\frac{z^4}{R}\nonumber\\
    &+\mathcal{O}(z^6).
\end{align}
Notice that for $D=d$, this reduces to the series expansion of the semicircle $\sqrt{R^2-z^2}$. Since the solution is only known perturbatively, the coefficient of $z^D$ is ill-defined\footnote{This can be seen from the fact that the coefficient of $z^D$ would na\"ively have a $1/(D-D)$ prefactor.} and should be absorbed into $r_D$. As before, we write the general surface profile as
\begin{equation}
    r(z)=r_0(z)+r_D(R) z^D+\mathcal{O}(z^{D+2\Delta_+}).
\end{equation}

Now, we may compute the expansion of $v$ to be
\begin{equation}
    v(z)=1+\qty(\frac{1}{R}\dv{}{R}\qty(R r_D)z^D+o\qty(z^D))+\qty(\frac{(d-2)^2(D-d)}{2(D-2)^3(D-4)}\frac{z^4}{R^4}+o\qty(z^{4})).
\end{equation}
The $z^4$ term is potentially problematic and notably does not exist for the case $D=d$, which is due to the fact that
\begin{equation}
    \frac{\sqrt{R^2-z^2}}{R}\dv{}{R}\sqrt{R^2-z^2}=1.
\end{equation}
So there are three cases to consider:
\begin{enumerate}
    \item If $D<4$, then the $r_D$ term is still the leading order contribution, but since we assumed $D>d>2$, this is an empty collection of cases.
    \item If $D=4$, then we must have $d=3$. Moreover, the $z^4$ term arises from the $z^4$ term in $r_0$, which, as mentioned before, is ill-defined in $D=4$. Hence, the leading order contribution is the $r_4$ term as before, and the Lemma forces it to be positive
\begin{equation}
    \frac{1}{R}\dv{}{R}\qty(R r_4)\ge 0.
\end{equation}
Likewise, the derivative of the central charge is given by
\begin{equation}
    \dv{\Delta S'}{R}=-4\mathrm{Vol}(S^1)\mathrm{Vol}(M_1)\frac{1}{R}\dv{}{R}\qty(R r_4)+o(\epsilon^0).
\end{equation}
This proves the monotonicity of flows from $\mathrm{AdS}_5$ to $\mathrm{AdS}_4$.
    \item If $D>4$, then the leading term is the $z^4$ term, and the Lemma merely implies the tautology
\begin{equation}
    \frac{(d-2)^2(D-d)}{2(D-2)^3(D-4)}\frac{1}{R^4}\ge 0.
\end{equation}
However, the central charge derivative is given by
\begin{equation}
    \dv{}{R}\qty(\frac{\Delta S'}{R^{d-3}})=-D\mathrm{Vol}(S^{d-2})\mathrm{Vol}(M_{D-d})\frac{1}{R}\dv{}{R}\qty(Rr_D)+o(\epsilon^0).
\end{equation}
Unfortunately, the sign is no longer constrained by the Lemma, and so it becomes difficult to prove with our present approach. 
\end{enumerate}

\section{Discussion}\label{sec:discussion}
Our proof of irreversibility in RG flows contains explicitly several ingredients that could ultimately clarify the two main approaches to establishing such theorems: field-theoretic and entropic. Namely, in the expansion of metric terms for a given background solution, subleading terms in the Fefferman-Graham expansion encode information about the stress-energy tensor and its correlation functions \cite{Zamolodchikov:1986gt,Komargodski:2011vj}. We see how those terms affect the properties of the entanglement entropy, which we computed using the Ryu-Takayanagi prescription. We plan to pursue this lead to bridge the field theory proofs based on properties of correlators and amplitudes and the entropic proof based on strong subadditivity of the relative entropy \cite{Casini:2017vbe}.

One remaining question is how to generalize these results beyond $d=4$. Some progress has been made in this direction field theoretically in five dimensions \cite{Jafferis:2012iv,Chang:2017cdx,Fluder:2020pym} and six dimensions \cite{Elvang:2012st,Cordova:2015vwa,Cordova:2015fha,Heckman:2015axa,Mekareeya:2016yal,Heckman:2021nwg,Fazzi:2023ulb}, as well as holographically for Higgs branch flows in six dimensions \cite{Apruzzi:2017nck}. Of particular relevance to the present context, \cite{Liu:2012eea} proposed an interpolating $c$-function
\begin{equation}
    c_\mathrm{LM}(R)=\begin{cases}\frac{1}{(d-2)!!}(R\partial_R-1)(R\partial_R-3)\cdots(R\partial_R-(d-2))S_\mathrm{EE}(R)&d\text{ odd}\\
    \frac{1}{(d-2)!!}R\partial_R(R\partial_R-2)\cdots(R\partial_R-(d-2))S_\mathrm{EE}(R)&d\text{ even}\end{cases},
\end{equation}
which agrees with \eqref{eq:cCTT} for $d=2,3$ and generalizes to higher dimensions. Unfortunately, one issue with using $c_\mathrm{LM}$ is that it is unclear how to generalize $v(z)$, as such a function would have to contain higher-order $R$ derivatives of $r(z)$. A bigger issue, however, is that there is also no guarantee that $c_\mathrm{LM}$ is monotonic in general. Non-monotonic examples have been found in $d=4$, for instance. We hope to return to some of these issues elsewhere.

\section*{Acknowledgements}
We thank Jonathan Sorce and Antony Speranza for discussions. This work is partially supported by the U.S. Department of Energy under grant DE-SC0007859. ED and RJS were supported in part by Leinweber Graduate Summer Fellowships. RJS was supported in part by a Rackham One-Term Dissertation Fellowship.

\appendix
\section{Domain wall flows}\label{app:DWflows}
Here we discuss the corrections to the metric due to a scalar field. Consider a scalar field coupled to gravity
\begin{equation}
    e^{-1}\mathcal L=R-\fft12(\partial\phi)^2-V(\phi).
\end{equation}
Note that the scalar potential $V(\phi)$ implicitly includes the cosmological constant and also that the generalization to multiple scalars is straightforward. For a domain wall flow corresponding to the metric \eqref{eq:metric}, the coupled Einstein and Klein-Gordon equations are
\begin{align}
    &\phi''-(d-1)\fft{h'}h\phi'-\fft{\partial_\phi V}{h^2}=0,\nn\\
    &\fft{h''}h=\fft1{2(d-1)}(\phi')^2,\nn\\
    &\left(\fft{h'}h\right)^2=\fft1{2d(d-1)}(\phi')^2-\fft1{d(d-1)}\fft{V}{h^2}.    
\end{align}
One can check that the first and last equations imply the middle one.  For numerical solutions, one would typically solve the first two second-order equations and check that the last ``energy'' equation is satisfied.

Note that we can obtain a BPS flow by introducing a superpotential
\begin{equation}
    V=\fft12\left(\partial_\phi W^2-\fft{d}{2(d-1)}W^2\right).
\end{equation}
Then the first-order BPS flow is given by
\begin{equation}
    h'=\fft1{2(d-1)}W,\qquad\phi'=\fft1h\partial_\phi W.
\end{equation}

To understand the asymptotics, take a potential
\begin{equation}
    V=-\fft{d(d-1)}{L^2}+\fft12m^2\phi^2,
\end{equation}
corresponding to a cosmological constant and minimally coupled massive scalar. We now assume the asymptotics
\begin{equation}
    \phi\sim\alpha z^{\Delta},\qquad h\sim\fft{z}L\qty(1+\bar h(z)).
\end{equation}
The Klein-Gordon equation gives
\begin{equation}
    \Delta(\Delta-d)=(mL)^2\qquad\Rightarrow\qquad\Delta_\pm=\fft{d}2\pm\sqrt{\left(\fft{d}2\right)^2+(mL)^2}.
\end{equation}
One then interprets $\Delta_-$ as the 
conformal dimension of the perturbing operator and $\Delta_+$ as the dimension of the coupling. The first backreaction correction to the metric is then
\begin{equation}
    \bar h(z)=\fft{\Delta_+}{4(d-1)(2\Delta_++1)}\alpha^2z^{2\Delta_+}.
\end{equation}
Note that this expansion assumes $\Delta_+>0$, or, equivalently, $\Delta_-<d$, corresponding to a relevant deformation. This justifies our choice \eqref{eq:metric-exp}.

For a minimally coupled scalar, the series solution can be developed to higher order.  The result is an expansion of $\phi(z)$ in odd powers of $z^{\Delta_+}$ and an expansion of $\bar h(z)$ in even powers of $z^{\Delta_+}$.

\section{$r_d$ for small surfaces}\label{app:rd}
Since we do not have good control over the whole entangling surface, it is generally quite difficult to write down an expression for $r_d$. However, when $R$ is small, \emph{i.e.}, when we are close to the UV, one can find an explicit expression, similar to what was done in the small $R$ expansion in \cite{Liu:2012eea}. Let us take $R$ to be small and
\begin{equation}
    x=\frac{z}{R}\in\qty[0,\frac{z_0}{R}],
\end{equation}
fixed. We will, as before, take the warp factor to have an asymptotic expansion
\begin{equation}
    h(z)=z+c z^{1+2\Delta_+}+o\qty(z^{1+2\Delta_+}).
\end{equation}
Then we may expand the equations of motion
\begin{equation}
    r''(x) = R^2\left(\frac{d-2}{r(x)} + (d-1)\frac{h'(x)r'(x)}{R^2h(x)}\right)\left(1+\frac{r'(x)^2}{R^2}\right),
\end{equation}
perturbatively in $R$ and solve order-by-order to get an expansion for the profile
\begin{align}
    r(x)&=R\sqrt{1-x^2}+\frac{R^{1+2\Delta_+}}{\sqrt{1-x^2}}\bigg\{c_1+c_2x^d\,{}_2 F_1\qty(\frac{d-1}{2},\frac{d}{2};\frac{d+2}{2};x^2)\nn\\
    &\quad+\frac{2\Delta_+ (d-1)cx^{2+2\Delta_+}}{2d(d-2-2\Delta_+}\bigg[(d-2-2\Delta_+)\frac{\Gamma\qty(\Delta_++1)}{\Gamma\qty(\Delta_++2)}{}_3F_2\qty(\frac{3}{2},1,1+\Delta_+;\frac{d+2}{2},2+\Delta_+;x^2)\nn\\
    &\quad+2\,{}_2F_1\qty(\frac{3-d}{2},\frac{2+2\Delta_+-d}{2};\frac{4+2\Delta_+-d}{2};x^2){}_2F_1\qty(\frac{d-1}{2},\frac{d}{2};\frac{d+2}{2};x^2)\bigg]\bigg\}\nn\\
    &\quad+o(R^{1+2\Delta_+}).
\end{align}
Fixing the turning point\footnote{Note that this allows $r(0)$ to change, but we may always rescale the surface in the end to set $r(0)=R$. This will not affect $r_d$ to leading order in $R$.} to be $z_0=R$ and requiring that the surface is well-behaved at $x=1$ fixes the coefficients $c_1$ and $c_2$, which in turn allows us to extract $r_d$ for various dimensions, which we have presented in Table \ref{tab:rd}. Note that these will receive corrections in $R$ as the surface grows.  One can also check that these all satisfy $\dd\qty(R r_d)/\dd R\ge 0$ to leading order in $R$, so long as $2\Delta_+\ge d-4$.

\begin{table}[t]
    \centering
    \begin{tabular}{|l|r|}
        \hline
        $d$ & $r_d(R)$ \\
        \hline
        2 & $\frac{c}{2}R^{-1+2\Delta_+}$\\
        3 & $\frac{4\Delta_+ c}{3(2\Delta_+-1)}R^{-2+2\Delta_+}$\\
        4& $\frac{3c}{16(\Delta_+-1)}\qty(\frac{8\sqrt{\pi}\Gamma\qty(\Delta_++1)}{\Gamma\qty(\Delta_++\frac{1}{2})}-6\qty(\Delta_++1))R^{-3+2\Delta_+}$\\
        5& $\frac{16\Delta_+ c}{5(4\Delta_+^2-8\Delta_++3)}R^{-4+2\Delta_+}$\\
        6& $\frac{5c}{192(\Delta_+-1)(\Delta_+-2)}\qty(120-20\Delta_+(\Delta_+-7)-\frac{96\sqrt{\pi}\Gamma\qty(\Delta_++1)}{\Gamma\qty(\Delta_++\frac{1}{2})})R^{-5+2\Delta_+}$\\
        \hline
    \end{tabular}
    \caption{Table of various values of $r_d$ for small surfaces, up to $o(R^{-d+2\Delta_+})$ corrections.}
    \label{tab:rd}
\end{table}

\bibliographystyle{JHEP}
\bibliography{IrreversibilityRefs.bib}

\providecommand{\href}[2]{#2}\begingroup\raggedright\begin{thebibliography}{10}

\bibitem{Zamolodchikov:1986gt}
A.B.~Zamolodchikov, \emph{{Irreversibility of the Flux of the Renormalization
  Group in a 2D Field Theory}}, {\emph{JETP Lett.} {\bfseries 43} (1986) 730}.

\bibitem{Komargodski:2011vj}
Z.~Komargodski and A.~Schwimmer, \emph{{On Renormalization Group Flows in Four
  Dimensions}}, \href{https://doi.org/10.1007/JHEP12(2011)099}{\emph{JHEP}
  {\bfseries 12} (2011) 099} [\href{https://arxiv.org/abs/1107.3987}{{\ttfamily
  1107.3987}}].

\bibitem{Elvang:2012st}
H.~Elvang, D.Z.~Freedman, L.-Y.~Hung, M.~Kiermaier, R.C.~Myers and S.~Theisen,
  \emph{{On renormalization group flows and the a-theorem in 6d}},
  \href{https://doi.org/10.1007/JHEP10(2012)011}{\emph{JHEP} {\bfseries 10}
  (2012) 011} [\href{https://arxiv.org/abs/1205.3994}{{\ttfamily 1205.3994}}].

\bibitem{Cordova:2015vwa}
C.~Cordova, T.T.~Dumitrescu and X.~Yin, \emph{{Higher derivative terms,
  toroidal compactification, and Weyl anomalies in six-dimensional (2, 0)
  theories}}, \href{https://doi.org/10.1007/JHEP10(2019)128}{\emph{JHEP}
  {\bfseries 10} (2019) 128}
  [\href{https://arxiv.org/abs/1505.03850}{{\ttfamily 1505.03850}}].

\bibitem{Cordova:2015fha}
C.~Cordova, T.T.~Dumitrescu and K.~Intriligator, \emph{{Anomalies,
  renormalization group flows, and the a-theorem in six-dimensional (1, 0)
  theories}}, \href{https://doi.org/10.1007/JHEP10(2016)080}{\emph{JHEP}
  {\bfseries 10} (2016) 080}
  [\href{https://arxiv.org/abs/1506.03807}{{\ttfamily 1506.03807}}].

\bibitem{Heckman:2021nwg}
J.J.~Heckman, S.~Kundu and H.Y.~Zhang, \emph{{Effective field theory of 6D SUSY
  RG Flows}}, \href{https://doi.org/10.1103/PhysRevD.104.085017}{\emph{Phys.
  Rev. D} {\bfseries 104} (2021) 085017}
  [\href{https://arxiv.org/abs/2103.13395}{{\ttfamily 2103.13395}}].

\bibitem{Casini:2004bw}
H.~Casini and M.~Huerta, \emph{{A Finite entanglement entropy and the
  c-theorem}},
  \href{https://doi.org/10.1016/j.physletb.2004.08.072}{\emph{Phys. Lett. B}
  {\bfseries 600} (2004) 142}
  [\href{https://arxiv.org/abs/hep-th/0405111}{{\ttfamily hep-th/0405111}}].

\bibitem{Casini:2006es}
H.~Casini and M.~Huerta, \emph{{A c-theorem for the entanglement entropy}},
  \href{https://doi.org/10.1088/1751-8113/40/25/S57}{\emph{J. Phys. A}
  {\bfseries 40} (2007) 7031}
  [\href{https://arxiv.org/abs/cond-mat/0610375}{{\ttfamily
  cond-mat/0610375}}].

\bibitem{Casini:2012ei}
H.~Casini and M.~Huerta, \emph{{On the RG running of the entanglement entropy
  of a circle}}, \href{https://doi.org/10.1103/PhysRevD.85.125016}{\emph{Phys.
  Rev. D} {\bfseries 85} (2012) 125016}
  [\href{https://arxiv.org/abs/1202.5650}{{\ttfamily 1202.5650}}].

\bibitem{Casini:2017vbe}
H.~Casini, E.~Test\'e and G.~Torroba, \emph{{Markov Property of the Conformal
  Field Theory Vacuum and the a Theorem}},
  \href{https://doi.org/10.1103/PhysRevLett.118.261602}{\emph{Phys. Rev. Lett.}
  {\bfseries 118} (2017) 261602}
  [\href{https://arxiv.org/abs/1704.01870}{{\ttfamily 1704.01870}}].

\bibitem{Kelly:2014mra}
W.R.~Kelly and A.C.~Wall, \emph{{Holographic proof of the averaged null energy
  condition}}, \href{https://doi.org/10.1103/PhysRevD.90.106003}{\emph{Phys.
  Rev. D} {\bfseries 90} (2014) 106003}
  [\href{https://arxiv.org/abs/1408.3566}{{\ttfamily 1408.3566}}].

\bibitem{Klinkhammer:1991ki}
G.~Klinkhammer, \emph{{Averaged energy conditions for free scalar fields in
  flat space-times}},
  \href{https://doi.org/10.1103/PhysRevD.43.2542}{\emph{Phys. Rev. D}
  {\bfseries 43} (1991) 2542}.

\bibitem{Faulkner:2016mzt}
T.~Faulkner, R.G.~Leigh, O.~Parrikar and H.~Wang, \emph{{Modular Hamiltonians
  for Deformed Half-Spaces and the Averaged Null Energy Condition}},
  \href{https://doi.org/10.1007/JHEP09(2016)038}{\emph{JHEP} {\bfseries 09}
  (2016) 038} [\href{https://arxiv.org/abs/1605.08072}{{\ttfamily
  1605.08072}}].

\bibitem{Hartman:2016lgu}
T.~Hartman, S.~Kundu and A.~Tajdini, \emph{{Averaged Null Energy Condition from
  Causality}}, \href{https://doi.org/10.1007/JHEP07(2017)066}{\emph{JHEP}
  {\bfseries 07} (2017) 066}
  [\href{https://arxiv.org/abs/1610.05308}{{\ttfamily 1610.05308}}].

\bibitem{Kravchuk:2018htv}
P.~Kravchuk and D.~Simmons-Duffin, \emph{{Light-ray operators in conformal
  field theory}}, \href{https://doi.org/10.1007/JHEP11(2018)102}{\emph{JHEP}
  {\bfseries 11} (2018) 102}
  [\href{https://arxiv.org/abs/1805.00098}{{\ttfamily 1805.00098}}].

\bibitem{Hofman:2008ar}
D.M.~Hofman and J.~Maldacena, \emph{{Conformal collider physics: Energy and
  charge correlations}},
  \href{https://doi.org/10.1088/1126-6708/2008/05/012}{\emph{JHEP} {\bfseries
  05} (2008) 012} [\href{https://arxiv.org/abs/0803.1467}{{\ttfamily
  0803.1467}}].

\bibitem{Hofman:2009ug}
D.M.~Hofman, \emph{{Higher Derivative Gravity, Causality and Positivity of
  Energy in a UV complete QFT}},
  \href{https://doi.org/10.1016/j.nuclphysb.2009.08.001}{\emph{Nucl. Phys. B}
  {\bfseries 823} (2009) 174}
  [\href{https://arxiv.org/abs/0907.1625}{{\ttfamily 0907.1625}}].

\bibitem{Hartman:2023qdn}
T.~Hartman and G.~Mathys, \emph{{Averaged null energy and the renormalization
  group}}, \href{https://doi.org/10.1007/JHEP12(2023)139}{\emph{JHEP}
  {\bfseries 12} (2023) 139}
  [\href{https://arxiv.org/abs/2309.14409}{{\ttfamily 2309.14409}}].

\bibitem{Hartman:2023ccw}
T.~Hartman and G.~Mathys, \emph{{Null energy constraints on two-dimensional RG
  flows}}, \href{https://doi.org/10.1007/JHEP01(2024)102}{\emph{JHEP}
  {\bfseries 01} (2024) 102}
  [\href{https://arxiv.org/abs/2310.15217}{{\ttfamily 2310.15217}}].

\bibitem{Casini:2016udt}
H.~Casini, E.~Teste and G.~Torroba, \emph{{Relative entropy and the RG flow}},
  \href{https://doi.org/10.1007/JHEP03(2017)089}{\emph{JHEP} {\bfseries 03}
  (2017) 089} [\href{https://arxiv.org/abs/1611.00016}{{\ttfamily
  1611.00016}}].

\bibitem{Bousso:2015mna}
R.~Bousso, Z.~Fisher, S.~Leichenauer and A.C.~Wall, \emph{{Quantum focusing
  conjecture}}, \href{https://doi.org/10.1103/PhysRevD.93.064044}{\emph{Phys.
  Rev. D} {\bfseries 93} (2016) 064044}
  [\href{https://arxiv.org/abs/1506.02669}{{\ttfamily 1506.02669}}].

\bibitem{Bousso:2015wca}
R.~Bousso, Z.~Fisher, J.~Koeller, S.~Leichenauer and A.C.~Wall, \emph{{Proof of
  the Quantum Null Energy Condition}},
  \href{https://doi.org/10.1103/PhysRevD.93.024017}{\emph{Phys. Rev. D}
  {\bfseries 93} (2016) 024017}
  [\href{https://arxiv.org/abs/1509.02542}{{\ttfamily 1509.02542}}].

\bibitem{Balakrishnan:2017bjg}
S.~Balakrishnan, T.~Faulkner, Z.U.~Khandker and H.~Wang, \emph{{A General Proof
  of the Quantum Null Energy Condition}},
  \href{https://doi.org/10.1007/JHEP09(2019)020}{\emph{JHEP} {\bfseries 09}
  (2019) 020} [\href{https://arxiv.org/abs/1706.09432}{{\ttfamily
  1706.09432}}].

\bibitem{Malik:2019dpg}
T.A.~Malik and R.~Lopez-Mobilia, \emph{{Proof of the quantum null energy
  condition for free fermionic field theories}},
  \href{https://doi.org/10.1103/PhysRevD.101.066028}{\emph{Phys. Rev. D}
  {\bfseries 101} (2020) 066028}
  [\href{https://arxiv.org/abs/1910.07594}{{\ttfamily 1910.07594}}].

\bibitem{Balakrishnan:2019gxl}
S.~Balakrishnan, V.~Chandrasekaran, T.~Faulkner, A.~Levine and
  A.~Shahbazi-Moghaddam, \emph{{Entropy variations and light ray operators from
  replica defects}}, \href{https://doi.org/10.1007/JHEP09(2022)217}{\emph{JHEP}
  {\bfseries 09} (2022) 217}
  [\href{https://arxiv.org/abs/1906.08274}{{\ttfamily 1906.08274}}].

\bibitem{Kudler-Flam:2023hkl}
J.~Kudler-Flam, S.~Leutheusser, A.A.~Rahman, G.~Satishchandran and
  A.J.~Speranza, \emph{{A covariant regulator for entanglement entropy: proofs
  of the Bekenstein bound and QNEC}},
  \href{https://arxiv.org/abs/2312.07646}{{\ttfamily 2312.07646}}.

\bibitem{Koeller:2015qmn}
J.~Koeller and S.~Leichenauer, \emph{{Holographic Proof of the Quantum Null
  Energy Condition}},
  \href{https://doi.org/10.1103/PhysRevD.94.024026}{\emph{Phys. Rev. D}
  {\bfseries 94} (2016) 024026}
  [\href{https://arxiv.org/abs/1512.06109}{{\ttfamily 1512.06109}}].

\bibitem{Leichenauer:2018obf}
S.~Leichenauer, A.~Levine and A.~Shahbazi-Moghaddam, \emph{{Energy density from
  second shape variations of the von Neumann entropy}},
  \href{https://doi.org/10.1103/PhysRevD.98.086013}{\emph{Phys. Rev. D}
  {\bfseries 98} (2018) 086013}
  [\href{https://arxiv.org/abs/1802.02584}{{\ttfamily 1802.02584}}].

\bibitem{Ceyhan:2018zfg}
F.~Ceyhan and T.~Faulkner, \emph{{Recovering the QNEC from the ANEC}},
  \href{https://doi.org/10.1007/s00220-020-03751-y}{\emph{Commun. Math. Phys.}
  {\bfseries 377} (2020) 999}
  [\href{https://arxiv.org/abs/1812.04683}{{\ttfamily 1812.04683}}].

\bibitem{Ecker:2020gnw}
C.~Ecker, D.~Grumiller, H.~Soltanpanahi and P.~Stanzer, \emph{{QNEC2 in
  deformed holographic CFTs}},
  \href{https://doi.org/10.1007/JHEP03(2021)213}{\emph{JHEP} {\bfseries 03}
  (2021) 213} [\href{https://arxiv.org/abs/2007.10367}{{\ttfamily
  2007.10367}}].

\bibitem{Casini:2023kyj}
H.~Casini, I.~Salazar~Landea and G.~Torroba, \emph{{Irreversibility, QNEC, and
  defects}}, \href{https://doi.org/10.1007/JHEP07(2023)004}{\emph{JHEP}
  {\bfseries 07} (2023) 004}
  [\href{https://arxiv.org/abs/2303.16935}{{\ttfamily 2303.16935}}].

\bibitem{Girardello:1998pd}
L.~Girardello, M.~Petrini, M.~Porrati and A.~Zaffaroni, \emph{{Novel local CFT
  and exact results on perturbations of N=4 superYang Mills from AdS
  dynamics}}, \href{https://doi.org/10.1088/1126-6708/1998/12/022}{\emph{JHEP}
  {\bfseries 12} (1998) 022}
  [\href{https://arxiv.org/abs/hep-th/9810126}{{\ttfamily hep-th/9810126}}].

\bibitem{Freedman:1999gp}
D.Z.~Freedman, S.S.~Gubser, K.~Pilch and N.P.~Warner, \emph{{Renormalization
  group flows from holography supersymmetry and a c theorem}},
  \href{https://doi.org/10.4310/ATMP.1999.v3.n2.a7}{\emph{Adv. Theor. Math.
  Phys.} {\bfseries 3} (1999) 363}
  [\href{https://arxiv.org/abs/hep-th/9904017}{{\ttfamily hep-th/9904017}}].

\bibitem{Myers:2010tj}
R.C.~Myers and A.~Sinha, \emph{{Holographic c-theorems in arbitrary
  dimensions}}, \href{https://doi.org/10.1007/JHEP01(2011)125}{\emph{JHEP}
  {\bfseries 01} (2011) 125} [\href{https://arxiv.org/abs/1011.5819}{{\ttfamily
  1011.5819}}].

\bibitem{Myers:2010xs}
R.C.~Myers and A.~Sinha, \emph{{Seeing a c-theorem with holography}},
  \href{https://doi.org/10.1103/PhysRevD.82.046006}{\emph{Phys. Rev. D}
  {\bfseries 82} (2010) 046006}
  [\href{https://arxiv.org/abs/1006.1263}{{\ttfamily 1006.1263}}].

\bibitem{Anber:2008js}
M.M.~Anber and D.~Kastor, \emph{{C-Functions in Lovelock Gravity}},
  \href{https://doi.org/10.1088/1126-6708/2008/05/061}{\emph{JHEP} {\bfseries
  05} (2008) 061} [\href{https://arxiv.org/abs/0802.1290}{{\ttfamily
  0802.1290}}].

\bibitem{Sinha:2010ai}
A.~Sinha, \emph{{On the new massive gravity and AdS/CFT}},
  \href{https://doi.org/10.1007/JHEP06(2010)061}{\emph{JHEP} {\bfseries 06}
  (2010) 061} [\href{https://arxiv.org/abs/1003.0683}{{\ttfamily 1003.0683}}].

\bibitem{Gullu:2010pc}
I.~Gullu, T.C.~Sisman and B.~Tekin, \emph{{Born-Infeld extension of new massive
  gravity}}, \href{https://doi.org/10.1088/0264-9381/27/16/162001}{\emph{Class.
  Quant. Grav.} {\bfseries 27} (2010) 162001}
  [\href{https://arxiv.org/abs/1003.3935}{{\ttfamily 1003.3935}}].

\bibitem{Gullu:2010st}
I.~Gullu, T.C.~Sisman and B.~Tekin, \emph{{c-functions in the Born-Infeld
  extended New Massive Gravity}},
  \href{https://doi.org/10.1103/PhysRevD.82.024032}{\emph{Phys. Rev. D}
  {\bfseries 82} (2010) 024032}
  [\href{https://arxiv.org/abs/1005.3214}{{\ttfamily 1005.3214}}].

\bibitem{Sinha:2010pm}
A.~Sinha, \emph{{On higher derivative gravity, $c$-theorems and cosmology}},
  \href{https://doi.org/10.1088/0264-9381/28/8/085002}{\emph{Class. Quant.
  Grav.} {\bfseries 28} (2011) 085002}
  [\href{https://arxiv.org/abs/1008.4315}{{\ttfamily 1008.4315}}].

\bibitem{Oliva:2010eb}
J.~Oliva and S.~Ray, \emph{{A new cubic theory of gravity in five dimensions:
  Black hole, Birkhoff's theorem and C-function}},
  \href{https://doi.org/10.1088/0264-9381/27/22/225002}{\emph{Class. Quant.
  Grav.} {\bfseries 27} (2010) 225002}
  [\href{https://arxiv.org/abs/1003.4773}{{\ttfamily 1003.4773}}].

\bibitem{Myers:2010ru}
R.C.~Myers and B.~Robinson, \emph{{Black Holes in Quasi-topological Gravity}},
  \href{https://doi.org/10.1007/JHEP08(2010)067}{\emph{JHEP} {\bfseries 08}
  (2010) 067} [\href{https://arxiv.org/abs/1003.5357}{{\ttfamily 1003.5357}}].

\bibitem{Oliva:2010zd}
J.~Oliva and S.~Ray, \emph{{Classification of Six Derivative Lagrangians of
  Gravity and Static Spherically Symmetric Solutions}},
  \href{https://doi.org/10.1103/PhysRevD.82.124030}{\emph{Phys. Rev. D}
  {\bfseries 82} (2010) 124030}
  [\href{https://arxiv.org/abs/1004.0737}{{\ttfamily 1004.0737}}].

\bibitem{Liu:2010xc}
J.T.~Liu, W.~Sabra and Z.~Zhao, \emph{{Holographic c-theorems and higher
  derivative gravity}},
  \href{https://doi.org/10.1103/PhysRevD.85.126004}{\emph{Phys. Rev. D}
  {\bfseries 85} (2012) 126004}
  [\href{https://arxiv.org/abs/1012.3382}{{\ttfamily 1012.3382}}].

\bibitem{Liu:2011iia}
J.T.~Liu and Z.~Zhao, \emph{{A holographic c-theorem for higher derivative
  gravity}},  \href{https://arxiv.org/abs/1108.5179}{{\ttfamily 1108.5179}}.

\bibitem{Alkac:2018whk}
G.~Alka\c{c} and B.~Tekin, \emph{{Holographic c-theorem and Born-Infeld Gravity
  Theories}}, \href{https://doi.org/10.1103/PhysRevD.98.046013}{\emph{Phys.
  Rev. D} {\bfseries 98} (2018) 046013}
  [\href{https://arxiv.org/abs/1805.07963}{{\ttfamily 1805.07963}}].

\bibitem{Ghodsi:2019xrx}
A.~Ghodsi and M.~Siahvoshan, \emph{{A Holographic Study of the $a$-theorem and
  RG Flow in General Quadratic Curvature Gravity}},
  \href{https://doi.org/10.1140/epjc/s10052-019-7345-8}{\emph{Eur. Phys. J. C}
  {\bfseries 79} (2019) 820}
  [\href{https://arxiv.org/abs/1907.03497}{{\ttfamily 1907.03497}}].

\bibitem{Anastasiou:2021jcv}
G.~Anastasiou, I.J.~Araya, R.B.~Mann and R.~Olea, \emph{{Renormalized
  holographic entanglement entropy in Lovelock gravity}},
  \href{https://doi.org/10.1007/JHEP06(2021)073}{\emph{JHEP} {\bfseries 06}
  (2021) 073} [\href{https://arxiv.org/abs/2103.14640}{{\ttfamily
  2103.14640}}].

\bibitem{Ghodsi:2021xrb}
A.~Ghodsi and M.~Siahvoshan, \emph{{Higher order curvature corrections and
  holographic renormalization group flow}},
  \href{https://doi.org/10.1103/PhysRevD.104.126025}{\emph{Phys. Rev. D}
  {\bfseries 104} (2021) 126025}
  [\href{https://arxiv.org/abs/2105.13208}{{\ttfamily 2105.13208}}].

\bibitem{Alkac:2022zda}
G.~Alkac and G.~Suer, \emph{{3D Lovelock gravity and the holographic
  c-theorem}}, \href{https://doi.org/10.1103/PhysRevD.107.046014}{\emph{Phys.
  Rev. D} {\bfseries 107} (2023) 046014}
  [\href{https://arxiv.org/abs/2211.12450}{{\ttfamily 2211.12450}}].

\bibitem{Hartman:2022njz}
T.~Hartman, Y.~Jiang, F.~Sgarlata and A.~Tajdini, \emph{{Focusing bounds for
  CFT correlators and the S-matrix}},
  \href{https://arxiv.org/abs/2212.01942}{{\ttfamily 2212.01942}}.

\bibitem{Casini:2011kv}
H.~Casini, M.~Huerta and R.C.~Myers, \emph{{Towards a derivation of holographic
  entanglement entropy}},
  \href{https://doi.org/10.1007/JHEP05(2011)036}{\emph{JHEP} {\bfseries 05}
  (2011) 036} [\href{https://arxiv.org/abs/1102.0440}{{\ttfamily 1102.0440}}].

\bibitem{Albash:2011nq}
T.~Albash and C.V.~Johnson, \emph{{Holographic Entanglement Entropy and
  Renormalization Group Flow}},
  \href{https://doi.org/10.1007/JHEP02(2012)095}{\emph{JHEP} {\bfseries 02}
  (2012) 095} [\href{https://arxiv.org/abs/1110.1074}{{\ttfamily 1110.1074}}].

\bibitem{Myers:2012ed}
R.C.~Myers and A.~Singh, \emph{{Comments on Holographic Entanglement Entropy
  and RG Flows}}, \href{https://doi.org/10.1007/JHEP04(2012)122}{\emph{JHEP}
  {\bfseries 04} (2012) 122} [\href{https://arxiv.org/abs/1202.2068}{{\ttfamily
  1202.2068}}].

\bibitem{Liu:2013una}
H.~Liu and M.~Mezei, \emph{{Probing renormalization group flows using
  entanglement entropy}},
  \href{https://doi.org/10.1007/JHEP01(2014)098}{\emph{JHEP} {\bfseries 01}
  (2014) 098} [\href{https://arxiv.org/abs/1309.6935}{{\ttfamily 1309.6935}}].

\bibitem{Casini:2015ffa}
H.~Casini, E.~Teste and G.~Torroba, \emph{{Holographic RG flows, entanglement
  entropy and the sum rule}},
  \href{https://doi.org/10.1007/JHEP03(2016)033}{\emph{JHEP} {\bfseries 03}
  (2016) 033} [\href{https://arxiv.org/abs/1510.02103}{{\ttfamily
  1510.02103}}].

\bibitem{Park:2018ebm}
C.~Park, D.~Ro and J.~Hun~Lee, \emph{{c-theorem of the entanglement entropy}},
  \href{https://doi.org/10.1007/JHEP11(2018)165}{\emph{JHEP} {\bfseries 11}
  (2018) 165} [\href{https://arxiv.org/abs/1806.09072}{{\ttfamily
  1806.09072}}].

\bibitem{Daguerre:2022uxt}
L.~Daguerre, M.~Ginzburg and G.~Torroba, \emph{{Holographic entanglement
  entropy inequalities beyond strong subadditivity}},
  \href{https://doi.org/10.1007/JHEP10(2022)199}{\emph{JHEP} {\bfseries 10}
  (2022) 199} [\href{https://arxiv.org/abs/2208.03334}{{\ttfamily
  2208.03334}}].

\bibitem{Wall:2012uf}
A.C.~Wall, \emph{{Maximin Surfaces, and the Strong Subadditivity of the
  Covariant Holographic Entanglement Entropy}},
  \href{https://doi.org/10.1088/0264-9381/31/22/225007}{\emph{Class. Quant.
  Grav.} {\bfseries 31} (2014) 225007}
  [\href{https://arxiv.org/abs/1211.3494}{{\ttfamily 1211.3494}}].

\bibitem{Ryu:2006bv}
S.~Ryu and T.~Takayanagi, \emph{{Holographic derivation of entanglement entropy
  from AdS/CFT}},
  \href{https://doi.org/10.1103/PhysRevLett.96.181602}{\emph{Phys. Rev. Lett.}
  {\bfseries 96} (2006) 181602}
  [\href{https://arxiv.org/abs/hep-th/0603001}{{\ttfamily hep-th/0603001}}].

\bibitem{Ryu:2006ef}
S.~Ryu and T.~Takayanagi, \emph{{Aspects of Holographic Entanglement Entropy}},
  \href{https://doi.org/10.1088/1126-6708/2006/08/045}{\emph{JHEP} {\bfseries
  08} (2006) 045} [\href{https://arxiv.org/abs/hep-th/0605073}{{\ttfamily
  hep-th/0605073}}].

\bibitem{Maldacena:2000mw}
J.M.~Maldacena and C.~Nunez, \emph{{Supergravity description of field theories
  on curved manifolds and a no go theorem}},
  \href{https://doi.org/10.1142/S0217751X01003937}{\emph{Int. J. Mod. Phys. A}
  {\bfseries 16} (2001) 822}
  [\href{https://arxiv.org/abs/hep-th/0007018}{{\ttfamily hep-th/0007018}}].

\bibitem{Acharya:2000mu}
B.S.~Acharya, J.P.~Gauntlett and N.~Kim, \emph{{Five-branes wrapped on
  associative three cycles}},
  \href{https://doi.org/10.1103/PhysRevD.63.106003}{\emph{Phys. Rev. D}
  {\bfseries 63} (2001) 106003}
  [\href{https://arxiv.org/abs/hep-th/0011190}{{\ttfamily hep-th/0011190}}].

\bibitem{Gauntlett:2000ng}
J.P.~Gauntlett, N.~Kim and D.~Waldram, \emph{{M Five-branes wrapped on
  supersymmetric cycles}},
  \href{https://doi.org/10.1103/PhysRevD.63.126001}{\emph{Phys. Rev. D}
  {\bfseries 63} (2001) 126001}
  [\href{https://arxiv.org/abs/hep-th/0012195}{{\ttfamily hep-th/0012195}}].

\bibitem{Gauntlett:2001jj}
J.P.~Gauntlett and N.~Kim, \emph{{M five-branes wrapped on supersymmetric
  cycles. 2.}}, \href{https://doi.org/10.1103/PhysRevD.65.086003}{\emph{Phys.
  Rev. D} {\bfseries 65} (2002) 086003}
  [\href{https://arxiv.org/abs/hep-th/0109039}{{\ttfamily hep-th/0109039}}].

\bibitem{Gauntlett:2001qs}
J.P.~Gauntlett, N.~Kim, S.~Pakis and D.~Waldram, \emph{{Membranes wrapped on
  holomorphic curves}},
  \href{https://doi.org/10.1103/PhysRevD.65.026003}{\emph{Phys. Rev. D}
  {\bfseries 65} (2002) 026003}
  [\href{https://arxiv.org/abs/hep-th/0105250}{{\ttfamily hep-th/0105250}}].

\bibitem{Benini:2013cda}
F.~Benini and N.~Bobev, \emph{{Two-dimensional SCFTs from wrapped branes and
  c-extremization}}, \href{https://doi.org/10.1007/JHEP06(2013)005}{\emph{JHEP}
  {\bfseries 06} (2013) 005} [\href{https://arxiv.org/abs/1302.4451}{{\ttfamily
  1302.4451}}].

\bibitem{Benini:2015bwz}
F.~Benini, N.~Bobev and P.M.~Crichigno, \emph{{Two-dimensional SCFTs from
  D3-branes}}, \href{https://doi.org/10.1007/JHEP07(2016)020}{\emph{JHEP}
  {\bfseries 07} (2016) 020}
  [\href{https://arxiv.org/abs/1511.09462}{{\ttfamily 1511.09462}}].

\bibitem{Bobev:2017uzs}
N.~Bobev and P.M.~Crichigno, \emph{{Universal RG Flows Across Dimensions and
  Holography}}, \href{https://doi.org/10.1007/JHEP12(2017)065}{\emph{JHEP}
  {\bfseries 12} (2017) 065}
  [\href{https://arxiv.org/abs/1708.05052}{{\ttfamily 1708.05052}}].

\bibitem{Macpherson:2014eza}
N.T.~Macpherson, C.~N\'u\~nez, L.A.~Pando~Zayas, V.G.J.~Rodgers and
  C.A.~Whiting, \emph{{Type IIB supergravity solutions with AdS$_{5}$ from
  Abelian and non-Abelian T dualities}},
  \href{https://doi.org/10.1007/JHEP02(2015)040}{\emph{JHEP} {\bfseries 02}
  (2015) 040} [\href{https://arxiv.org/abs/1410.2650}{{\ttfamily 1410.2650}}].

\bibitem{Bea:2015fja}
Y.~Bea, J.D.~Edelstein, G.~Itsios, K.S.~Kooner, C.~Nunez, D.~Schofield et~al.,
  \emph{{Compactifications of the Klebanov-Witten CFT and new AdS$_{3}$
  backgrounds}}, \href{https://doi.org/10.1007/JHEP05(2015)062}{\emph{JHEP}
  {\bfseries 05} (2015) 062}
  [\href{https://arxiv.org/abs/1503.07527}{{\ttfamily 1503.07527}}].

\bibitem{Legramandi:2021aqv}
A.~Legramandi and C.~Nunez, \emph{{Holographic description of SCFT$_{5}$
  compactifications}},
  \href{https://doi.org/10.1007/JHEP02(2022)010}{\emph{JHEP} {\bfseries 02}
  (2022) 010} [\href{https://arxiv.org/abs/2109.11554}{{\ttfamily
  2109.11554}}].

\bibitem{GonzalezLezcano:2022mcd}
A.~Gonz\'alez~Lezcano, J.~Hong, J.T.~Liu, L.A.~Pando~Zayas and C.F.~Uhlemann,
  \emph{{c-functions in flows across dimensions}},
  \href{https://doi.org/10.1007/JHEP10(2022)083}{\emph{JHEP} {\bfseries 10}
  (2022) 083} [\href{https://arxiv.org/abs/2207.09360}{{\ttfamily
  2207.09360}}].

\bibitem{Deddo:2023pid}
E.~Deddo, J.T.~Liu, L.A.~Pando~Zayas and R.J.~Saskowski, \emph{{c-functions in
  higher-derivative flows across dimensions}},
  \href{https://doi.org/10.1007/JHEP08(2023)147}{\emph{JHEP} {\bfseries 08}
  (2023) 147} [\href{https://arxiv.org/abs/2305.18530}{{\ttfamily
  2305.18530}}].

\bibitem{Deddo:2022wxj}
E.~Deddo, L.A.~Pando~Zayas and C.F.~Uhlemann, \emph{{Entanglement and topology
  in RG flows across dimensions: caps, bridges and corners}},
  \href{https://doi.org/10.1007/JHEP04(2023)018}{\emph{JHEP} {\bfseries 04}
  (2023) 018} [\href{https://arxiv.org/abs/2301.00257}{{\ttfamily
  2301.00257}}].

\bibitem{de-la-Cruz-Moreno:2023mew}
J.~de-la Cruz-Moreno, J.T.~Liu and L.A.~Pando~Zayas, \emph{{Discontinuity in RG
  Flows Across Dimensions: Entanglement, Anomaly Coefficients and Geometry}},
  \href{https://arxiv.org/abs/2312.12382}{{\ttfamily 2312.12382}}.

\bibitem{Maldacena:1997re}
J.M.~Maldacena, \emph{{The large N limit of superconformal field theories and
  supergravity}}, {\emph{Adv. Theor. Math. Phys.} {\bfseries 2} (1998) 231}
  [\href{https://arxiv.org/abs/hep-th/9711200}{{\ttfamily hep-th/9711200}}].

\bibitem{Witten:2018zxz}
E.~Witten, \emph{{APS Medal for Exceptional Achievement in Research: Invited
  article on entanglement properties of quantum field theory}},
  \href{https://doi.org/10.1103/RevModPhys.90.045003}{\emph{Rev. Mod. Phys.}
  {\bfseries 90} (2018) 045003}
  [\href{https://arxiv.org/abs/1803.04993}{{\ttfamily 1803.04993}}].

\bibitem{Speranza:2019hkr}
A.J.~Speranza, \emph{{Geometrical tools for embedding fields, submanifolds, and
  foliations}},  \href{https://arxiv.org/abs/1904.08012}{{\ttfamily
  1904.08012}}.

\bibitem{Engelhardt:2019hmr}
N.~Engelhardt and S.~Fischetti, \emph{{Surface Theory: the Classical, the
  Quantum, and the Holographic}},
  \href{https://doi.org/10.1088/1361-6382/ab3bda}{\emph{Class. Quant. Grav.}
  {\bfseries 36} (2019) 205002}
  [\href{https://arxiv.org/abs/1904.08423}{{\ttfamily 1904.08423}}].

\bibitem{Fefferman:1985}
C.~Fefferman and C.R.~Graham, \emph{Conformal invariants},  in \emph{\'Elie
  Cartan et les math\'ematiques d'aujourd'hui - Lyon, 25-29 juin 1984},
  no.~S131 in Ast\'erisque, Soci\'et\'e math\'ematique de France (1985),
  \href{http://www.numdam.org/item/AST\_1985\_\_S131\_\_95\_0/}{http://www.numdam.org/item/AST\_1985\_\_S131\_\_95\_0/}.

\bibitem{Kontou:2020bta}
E.-A.~Kontou and K.~Sanders, \emph{{Energy conditions in general relativity and
  quantum field theory}},
  \href{https://doi.org/10.1088/1361-6382/ab8fcf}{\emph{Class. Quant. Grav.}
  {\bfseries 37} (2020) 193001}
  [\href{https://arxiv.org/abs/2003.01815}{{\ttfamily 2003.01815}}].

\bibitem{Jafferis:2012iv}
D.L.~Jafferis and S.S.~Pufu, \emph{{Exact results for five-dimensional
  superconformal field theories with gravity duals}},
  \href{https://doi.org/10.1007/JHEP05(2014)032}{\emph{JHEP} {\bfseries 05}
  (2014) 032} [\href{https://arxiv.org/abs/1207.4359}{{\ttfamily 1207.4359}}].

\bibitem{Chang:2017cdx}
C.-M.~Chang, M.~Fluder, Y.-H.~Lin and Y.~Wang, \emph{{Spheres, Charges,
  Instantons, and Bootstrap: A Five-Dimensional Odyssey}},
  \href{https://doi.org/10.1007/JHEP03(2018)123}{\emph{JHEP} {\bfseries 03}
  (2018) 123} [\href{https://arxiv.org/abs/1710.08418}{{\ttfamily
  1710.08418}}].

\bibitem{Fluder:2020pym}
M.~Fluder and C.F.~Uhlemann, \emph{{Evidence for a 5d F-theorem}},
  \href{https://doi.org/10.1007/JHEP02(2021)192}{\emph{JHEP} {\bfseries 02}
  (2021) 192} [\href{https://arxiv.org/abs/2011.00006}{{\ttfamily
  2011.00006}}].

\bibitem{Heckman:2015axa}
J.J.~Heckman and T.~Rudelius, \emph{{Evidence for C-theorems in 6D SCFTs}},
  \href{https://doi.org/10.1007/JHEP09(2015)218}{\emph{JHEP} {\bfseries 09}
  (2015) 218} [\href{https://arxiv.org/abs/1506.06753}{{\ttfamily
  1506.06753}}].

\bibitem{Mekareeya:2016yal}
N.~Mekareeya, T.~Rudelius and A.~Tomasiello, \emph{{T-branes, Anomalies and
  Moduli Spaces in 6D SCFTs}},
  \href{https://doi.org/10.1007/JHEP10(2017)158}{\emph{JHEP} {\bfseries 10}
  (2017) 158} [\href{https://arxiv.org/abs/1612.06399}{{\ttfamily
  1612.06399}}].

\bibitem{Fazzi:2023ulb}
M.~Fazzi, S.~Giri and P.~Levy, \emph{{Proving the 6d a-theorem with the double
  affine Grassmannian}},  \href{https://arxiv.org/abs/2312.17178}{{\ttfamily
  2312.17178}}.

\bibitem{Apruzzi:2017nck}
F.~Apruzzi and M.~Fazzi, \emph{{AdS$_{7}$/CFT$_{6}$ with orientifolds}},
  \href{https://doi.org/10.1007/JHEP01(2018)124}{\emph{JHEP} {\bfseries 01}
  (2018) 124} [\href{https://arxiv.org/abs/1712.03235}{{\ttfamily
  1712.03235}}].

\bibitem{Liu:2012eea}
H.~Liu and M.~Mezei, \emph{{A Refinement of entanglement entropy and the number
  of degrees of freedom}},
  \href{https://doi.org/10.1007/JHEP04(2013)162}{\emph{JHEP} {\bfseries 04}
  (2013) 162} [\href{https://arxiv.org/abs/1202.2070}{{\ttfamily 1202.2070}}].

\end{thebibliography}\endgroup

\end{document}